\documentclass[a4paper]{article}
\pdfoutput=1
\usepackage{jcappub}
\usepackage{amsmath}
\usepackage{bm}
\usepackage{graphicx}
\usepackage{enumitem}
\usepackage{geometry}
\usepackage{xspace}
\usepackage{graphicx}
\usepackage{hyperref}
\usepackage{verbatim}
\usepackage{xcolor}

\allowdisplaybreaks 

\makeatletter
\gdef\@fpheader{}
\g@addto@macro\bfseries{\boldmath}
\makeatother

\newcommand{\deflen}[2]{%
    \expandafter\newlength\csname #1\endcsname
    \expandafter\setlength\csname #1\endcsname{#2}%
}

\usepackage{caption}
\usepackage{subcaption}
\usepackage{mathrsfs}
\usepackage{graphicx}

\newcommand{\exd}{{ \mathrm{d}} }
\def\be{\begin{equation}}
\def\ee{\end{equation}}
\def\bea{\begin{eqnarray}}
\def\eea{\end{eqnarray}}

\def\cA{{\mathcal{A}}}
\def\cB{{\mathcal{B}}}
\def\cC{{\mathcal{C}}}
\def\cO{{\mathcal{O}}}
\def\cD{{\mathcal{D}}}
\def\cE{{\mathcal{E}}}

\def\cI{{\cal I}}
\def\cJ{{\mathcal{J}}}

\def\cR{{\mathcal{R}}}
\def\cS{{\mathcal{S}}}
\def\cT{{\mathcal{T}}}

\def\cW{{\cal W}}
\def\cX{{\cal X}}
\def\cY{{\cal Y}}
\def\cZ{{\cal Z}}

\def\ssB{{\scriptscriptstyle B}}

\def\ssD{{\scriptscriptstyle D}}
\def\ssE{{\scriptscriptstyle E}}

\def\ssI{{\scriptscriptstyle I}}

\def\ssR{{\scriptscriptstyle R}}
\def\ssS{{\scriptscriptstyle \mathrm{S}}}
\def\ssT{{\scriptscriptstyle \mathrm{T}}}

\def\mfa{{\mathfrak a}}
\def\mfc{{\mathfrak c}}
\def\mfh{{\mathfrak h}}
\def\mfl{{\mathfrak l}}
\def\mfp{{\mathfrak p}}

\def\mfz{{\mathfrak{z}}}
\def\mfF{{\mathfrak{F}}}

\def\bmk{{\bm{k}}}
\def\bml{{\bm{l}}}

\def\bmq{{\bm{q}}}

\def\bmy{{\bm{y}}}

\def\scrC{\mathscr{C}}
\def\scrE{\mathscr{E}}
\def\scrH{\mathscr{H}}
\def\scrL{\mathscr{L}}
\def\scrW{\mathscr{W}}
\def\scrX{\mathscr{X}}
\def\scrY{\mathscr{Y}}
\def\scrZ{\mathscr{Z}}

\def\kUV{{k_{\UV}}}

\def\Tr{\mathrm{Tr}}

\def\Trenv{\underset{(e)}{\mathrm{Tr}}}
\def\smath#1{\text{\scalebox{.85}{$#1$}}}
\def\sfrac#1#2{\smath{\frac{#1}{#2}}}

\def\pref#1{(\ref{#1})}
\def\Mp{M_p}

\def\slrl{\epsilon_1}
\def\IR{{\scriptscriptstyle \mathrm{IR}}}
\def\UV{{\scriptscriptstyle \mathrm{UV}}}
\def\nn{\nonumber}
\newcommand{\roughly}[1]{\mathrel{\raise.3ex\hbox{$#1$\kern-0.75em
\lower1ex\hbox{$\sim$}}}}
\newcommand{\lsim}{\roughly<}

\def\w{\tilde v}
\def\p{\tilde p}
\def\sys{{(s)}}
\def\env{{(e)}}
\def\in{\mathrm{in}}

\subheader{}

\title{Inflationary Decoherence\\ from the Gravitational Floor}

\author[a,b,c]{C.P.~Burgess,}
\author[d]{R.~Holman}
\author[e]{and Greg Kaplanek}

\affiliation[a]{Department of Physics \& Astronomy,  McMaster University, Hamilton, ON, Canada
}

\affiliation[b]{Perimeter Institute for Theoretical Physics, Waterloo, ON, Canada
}
 
\affiliation[c]{School of Theoretical Physics, Dublin Institute for Advanced Studies, 
Dublin, Ireland}

\affiliation[d]{Minerva University, 14 Mint Plaza, San Francisco, CA, 
USA}

\affiliation[e]{Department of Electrical Engineering and Computer Science, Syracuse University, NY, USA}

\emailAdd{cburgess@perimeterinstitute.ca}
\emailAdd{rfcholman@gmail.com}
\emailAdd{gkaplane@syr.edu}

\date{today}

\begin{document}

\sloppy

\abstract{We re-examine the decoherence rate of primordial fluctuations within minimal inflationary models, using only the gravitational interactions required for the underlying fluctuation-generation mechanism itself. Since gravity provides the weakest interactions the result provides a plausible floor on the rate of primordial decoherence. Previous calculations (\href{https://arxiv.org/abs/2211.11046}{{\tt arXiv:2211.11046}}) did so using only a subset of these interactions, motivated by assuming both system and environment were super-Hubble. We extend this by including the effects on super-Hubble modes of {\it all} gravitational interactions amongst scalar fluctuations at leading order in $H/\Mp$ (and so need not restrict the decohering environment to being super-Hubble). We show how the decohering evolution becomes Markovian for super-Hubble modes, without the need to appeal to truncations (like the `rotating wave' approximation) that are often used in optics but can be inapprorpriate for cosmology. We find that the dominant contribution comes from the nonlocal cubic interactions obtained by solving the constraints. We find a decoherence rate that grows in the super-Hubble regime {\it faster} than found earlier and identify its leading divergent and finite parts. We argue why the divergent parts must cancel with other competing contributions -- such as decoherence due to environmental tensor modes -- that are partially computed elsewhere (and for which a complete calculation is in progress). We discuss the steps required to resum this result to late times and briefly discuss why they are more complicated than for earlier calculations. 
}

\maketitle

\section{Introduction}
\label{sec:intro}

Large-scale correlations in the distribution of matter and radiation throughout the observable universe have been observed in some detail \cite{eBOSS:2020yzd, Planck:2018vyg}, and their evolution is explained well by the standard $\Lambda$CDM cosmology provided they are seeded by a particular pattern of nearly scale-invariant primordial fluctuations inherited from the much-earlier universe. Fairly little is known about the origins of these primordial fluctuations, but their spectral properties famously agree with what would be expected of the quantum fluctuations of a potential-dominated gravitating scalar field, if these were to mix significantly with the metric and be stretched across the sky in the remote past by inflationary universal expansion \cite{Mukhanov:1981xt,Guth:1982ec,Hawking:1982cz,Starobinsky:1982ee,Bardeen:1983qw,Mukhanov:1988jd}.

Calculations of primordial decoherence \cite{Sakagami:1987mp, Brandenberger:1990bx, Matacz:1992mk, Lombardo:1995fg, Calzetta:1995ys, Polarski:1995jg, Kiefer:1998qe, Lombardo:2005iz, Burgess:2006jn, Prokopec:2006fc, Sharman:2007gi, Kiefer:2008ku, Franco:2011fg, Burgess:2014eoa, Nelson:2016kjm, Brahma:2020zpk, DaddiHammou:2022itk, Burgess:2022nwu, Colas:2022kfu, Sou:2022nsd, Boutivas:2023mfg, Colas:2024xjy, Colas:2024ysu, deKruijf:2024ufs, Sano:2025ird, Lopez:2025arw, Takeda:2025cye} try to quantify whether -- and how rapidly -- such early quantum fluctuations can decohere (before their detection) due to their interactions with other unobserved degrees of freedom. Decoherence allows initially quantum fluctuations to present as classical fluctuations in the much later universe probed by observations. Early-universe decoherence, if present, can obstruct efforts to find observational evidence for a quantum origin for primordial fluctuations \cite{Martin:2015qta,Martin:2017zxs,Campo:2005sv,Maldacena:2015bha,Martin:2021znx,Piotrak:2025zhy}. If decoherence is both early and rapid then quantum effects are unlikely to survive to the present day to be seen.  Reliable calculations of the dependence of the decoherence rate on the parameters of inflationary models are prerequisites for interpreting any such tests. 

Decoherence rates depend on precisely how strongly the observed degrees of freedom -- the `system' -- couple to any unmeasured sectors -- the `environment' -- with stronger couplings decohering more rapidly. A minimal assumption computes the decoherence rate using only the gravitational self-interactions of General Relativity (GR) that must in any case be present in any successful inflationary model. Since gravity is the weakest force its decoherence effects are likely to be as slow as is possible. Inclusion of other interactions likely only speeds up the decoherence process.

For gravitational calculations we work within the joint slow-roll and semi-classical expansions that typically control calculations in inflationary models. The small loop-counting parameter of semiclassical cosmology \cite{Burgess:2003jk, Adshead:2017srh} is $GH^2 = H^2/(8\pi \Mp^2)$, where $G$ is Newton's constant, $H$ is the inflationary Hubble scale and $\Mp$ is the reduced Planck mass. For near-inflationary geometries we also expand in powers of small slow-roll parameters, $\epsilon_i$, with the first of these parameterizing how slowly $H$ varies: $\slrl = - \dot H/H^2$ (while $\epsilon_2$ is proportional to $\dot \slrl/H$ and so on). For super-Hubble modes of co-moving momentum $\bm{k}$ we can also expand in powers of the small parameter $k/(a H)$ where $k = |\bm{k}|$. We argue in particular that non-Markovian evolution for super-Hubble modes is suppressed relative to Markovian evolution by powers of $k/(aH)$.

A preliminary study of how gravitational interactions decohere was made in \cite{Nelson:2016kjm}, who computed the evolution of the vacuum state for the inflaton's long-wavelength components once short wavelength modes are traced out, computing at leading order of perturbation theory using a specific representative of the various gravitational self-interactions predicted by GR. The computed rate diverged in the UV and suffered from `secular growth' (the generic breakdown of perturbative predictions at late times), making it hard to interpret reliably. 

Both of these issues were partially addressed in \cite{Burgess:2022nwu}, which computed how short-wavelength scalar modes decohere longer-wavelength tensor {\it and} scalar modes\footnote{Ref.~\cite{Burgess:2022nwu} also computes a contribution to the rate with which short-wavelength tensor modes decohere long-wavelength scalar modes, and a more complete calculation of this rate is in progress.} under the assumption that the observed `system' modes and the unseen `environment' modes are both super-Hubble. The system modes relevant to observations are indeed super-Hubble for most of their evolution and as it turns out this is what makes their evolution become Markovian. This Markovianity in turn allows the secular growth effects to be resummed so that late-time behaviour can be reliably accessed. 

The restrictive choice made in \cite{Burgess:2022nwu} was the assumption that the environment also be super-Hubble. This was taken on grounds of simplicity in order to justify the neglect of those gravitational interactions involving time derivatives, leaving a unique candidate interaction at lowest nontrivial order (the same interaction used in \cite{Nelson:2016kjm}). A drawback of this choice was that it becomes impossible to prepare the initial state (in which environment and system are uncorrelated) in the remote past, since in the remote past environmental modes are sub-Hubble in scale.

In this paper we extend the analysis of \cite{Burgess:2022nwu}, doing so mainly in two ways. First, we include the effects of {\it all} gravitational interactions among scalar fluctuations at leading order in the semiclassical and slow-roll expansions and drop the restriction to spatial derivatives. Doing so allows the analysis to apply to environmental modes that are sub-Hubble and this allows us to make a controlled adiabatic calculation starting with the Bunch-Davies state. We show that these new interactions not only can compete with the ones previously considered, they often dominate in the sense that they decohere more quickly (though a definitive statement requires also computing the contribution of tensor modes to scalar-mode decoherence). 

The second extension concerns the treatment of UV divergences. The regularization of UV divergences used in \cite{Burgess:2022nwu} was insensitive to power-law divergences (as would be dimensional regularization) but showed that logarithmic divergences had the form required to be absorbed into counterterms in the way expected by general arguments \cite{Burgess:2003jk, Adshead:2017srh}. They in particular dropped out of the leading decoherence result. We here regulate UV effects more systematically, keeping track of both power-law and logarithmic divergences, and find that some divergences remain. We expect these to be cancelled by as-yet uncomputed contributions, leaving a finite leading contribution, and this is shown more explicitly in \cite{ToAppear}. 

The endpoint of our analysis is a major contribution to the decoherence rate of super-Hubble modes -- eq.~\pref{LindbladPurity5} -- that differs from (and is larger than) the result found in \cite{Burgess:2022nwu}. We rederive the Markovian nature of evolution for super-Hubble system modes from first principles, showing it to be automatic at leading order in $k/(aH)$ for super-Hubble modes of wave-number $k$. Our calculation extends the derivation beyond the next-to-leading order (NLO) effects considered in \cite{Burgess:2022nwu}. Doing so reveals in particular how this affects the earlier calculations and ultimately changes the late-time behaviour. We find the dominant universal evolution due to gravitational interactions of scalar modes comes from the nonlocal cubic interaction in the classification of \cite{Maldacena:2002vr}, showing the important role played by the constraints. This is suggestive in light of the role played by constraints in the recent decoherence calculations of \cite{Danielson:2021egj, Danielson:2022tdw, Danielson:2022sga}.

\subsection{A Calculational Road Map}

Because the basic setup is very similar to previous calculations, for brevity's sake in what follows we refer the reader to \cite{Burgess:2022nwu} for many of the details and focus here on the things that change relative to the calculations described there. Our focus is again on the evolution of the reduced density matrix, $\langle \varphi_1 \, | \, \varrho \, | \, \varphi_2 \rangle$, of scalar perturbations, computed in the field basis, where 
\be 
   \varrho = \hbox{Tr}_{\rm env}\, \rho
\ee
is obtained by tracing out the environment of unobserved shorter-wavelength modes. 

The evolution of decoherence is found by computing the time-dependence of the purity 
\be \label{puritydef}
   \gamma := \hbox{Tr}_{\rm sys} \left( \varrho^2 \right) \,; 
\ee
a quantity that takes values between 0 and 1 and equals unity if and only if $\varrho$ describes a pure state.  We find $\gamma$ tends to zero at late times (for super-Hubble modes during inflation) and the goal is to compute the rate for this, taking care to do so without leaving the domain of validity of perturbation theory (which naively generically fails once $\gamma - 1$ is not perturbatively small). This decoherence occurs because $\varrho$ becomes diagonal in the field basis, and once this occurs it is indistinguishable from a non-quantum statistical ensemble of classical field configurations with probability distribution $P[\varphi] = \langle \varphi \, | \, \varrho \, | \, \varphi \rangle$. (It is the evolution of these diagonal probabilities that stochastic \cite{Starobinsky:1986fx,Starobinsky:1994bd,Mijic:1994vv,Seery:2010kh,Prokopec:2008gw} and de Sitter EFT \cite{Cohen:2020php, Cohen:2021fzf, Baumgart:2019clc} methods aim to compute.\footnote{As has been remarked elsewhere \cite{Kiefer:2006je, Burgess:2014eoa} the squeezing of modes during inflation \cite{Albrecht:1992kf} explains in a simple way why the density matrix diagonalizes in a basis of field eigenstates, making these the system's natural `pointer' basis.})

Our tools for computing the evolution of the off-diagonal density-matrix elements in perturbation theory build on earlier work that apply well-developed tools from the quantum theory of open systems to gravity, whose use to explore late-time evolution is known as Open Effective Field Theory (Open EFT)~\cite{Burgess:2014eoa, Agon:2014uxa, Burgess:2015ajz, Boyanovsky:2015xoa, Boyanovsky:2015tba, Boyanovsky:2015jen, Braaten:2016sja, Hollowood:2017bil, Martin:2018zbe, Martin:2018lin, Choudhury:2018ppd, Burrage:2018pyg, Kaplanek:2019dqu, Kaplanek:2019vzj, Jana:2020vyx, Loganayagam:2020eue,  Kaplanek:2020iay, Rai:2020edx, Banerjee:2021lqu,  Kaplanek:2021fnl, Burgess:2021luo, Colas:2022hlq, Brahma:2022yxu, brahma2022universal, DaddiHammou:2022itk,Kaplanek:2022xrr,Cao:2022kjn,Prudhoe:2022pte,Kading:2022jjl,Kading:2022hhc,Pelliconi:2023ojb,Kading:2023mdk,Keefe:2024cia,Brahma:2024yor,Bowen:2024emo,Bhattacharyya:2024duw,Salcedo:2024smn,Salcedo:2024nex,Tinwala:2024wod,Martin:2024mdm,Kranas:2025jgm,Salcedo:2025ezu}. These tools give the evolution equation for the reduced density matrix in the interaction picture as an approximate (schematic) form\footnote{We  use fundamental units throughout (for which $\hbar =  c = 1$).} 
\begin{equation}
\label{LindbladCartoon}
\partial_t \varrho = -i \left[ \overline\scrH_{\rm int} \,, \varrho \right]
+  \mathscr{L}_{2}\left(\varrho \right)+ \cO\left(\scrH_{\rm int}^3\right)
\,,
\end{equation}
where $\scrH_{\rm int}$ denotes the terms in the interaction Hamiltonian that couple the environment to the measured degrees of freedom and $\overline\scrH_{\rm int}$ denotes its average over the environment. The quantity $\mathscr{L}_2$, contains all of the contributions arising at second order in $\scrH_{\rm int}$, 

What is important is that some -- but not all -- of the terms in $\scrL_2$ can be written as the commutator of something with $\varrho$, what is often called `Hamiltonian evolution'.\footnote{We call this Hamiltonian evolution (as opposed to `unitarity', as is commonly used in the Open Systems literature) because in fundamental physics `unitarity' usually means the requirement that $\Tr\, \varrho = 1$ is preserved in time ({\it i.e.}~the sum of the probabilities for exhaustive and mutually exclusive alternatives always gives one). While it is true that unitarity {\it and} Hamiltonian evolution implies the evolution operator $U=e^{-iHt}$ is unitary (in the ordinary algebraic sense that $U^\star U = I$), for open systems Hamiltonian evolution need not imply the effective Hamiltonian is hermitian and non-hamiltonian evolution can preserve the condition $\Tr \varrho = 1$ (and so be unitary).} This is important for decoherence calculations because Hamiltonian evolution has the form of a Liouville equation for some choice of Hamiltonian -- for textbook reviews see \cite{BreuerPetruccione, Burgess:2020tbq} -- and so can never take pure states to mixed states. Because $\scrL_2$ is the leading contribution that cannot be expressed as Hamiltonian evolution it also provides the leading contribution to decoherence.

As argued in \cite{Burgess:2022nwu} this means the leading contributions to decoherence mediated by gravitational interactions arise at order $GH^2$ and come exclusively from the cubic interactions. Cubic interactions are central because they involve only one power of $1/\Mp$ and so are the least suppressed in a semiclassical approximation. But this means decoherence cannot be smaller than order $1/\Mp^2$ because it first arises at second order in eq.~\pref{LindbladCartoon}. Quartic and higher interactions are proportional to at least two powers of $1/\Mp$ and so can only appear at order $1/\Mp^2$ in the first-order term of \pref{LindbladCartoon} and so at this order cannot decohere. 

\subsubsection{The Lindblad limit (and its pitfalls)}

In some circumstances (as we show in detail below) $\mathscr{L}_2$ can be well-approximated by a Lindblad form~\cite{Lindblad:1975ef, Gorini:1976cm}, and when this occurs the evolution becomes Markovian in the sense that $\partial_t \varrho$ at time $t$ is completely determined by $\varrho$ also evaluated at time $t$ (rather than involving a convolution over its entire previous history). When the evolution is described by a Lindblad equation and when the coefficients in this Lindblad equation are not sensitive to the choice of the initial state then its solutions can be used to resum late-time evolution reliably even if the evolution equation itself is only computed perturbatively. This can be done because the evolution equation then has a broader domain of validity than does straight-up perturbation theory (along the lines of the program described more broadly in \cite{Burgess:2015ajz, Kaplanek:2019dqu, Kaplanek:2019vzj, Burgess:2022nwu, Burgess:2024eng} and reviewed in \cite{Burgess:2022rdo}). This makes a systematic understanding of the domain of validity of Markovian evolution important for making reliable predictions at late times. 

It is tempting when doing so to try to carry over techniques that have proven useful in other areas of physics for producing Markovian evolution. However the justification for these truncation techniques need not carry over to cosmology, for which the validity of Markovian evolution must be independently established (as was done in \cite{Burgess:2022nwu} and as we do here).  An example along these lines is the `rotating wave approximation' of quantum optics, whose validity when analyzing systems in the presence of a large laser signal, for instance, needn't apply in a cosmological context. 

Perturbation theory itself is sometimes proposed as the rationale for Markovian behaviour, on the grounds that non-Markovian behaviour arises when the right-hand side of \pref{LindbladCartoon} involves $\varrho(t')$ for $t' < t$ instead of just involving $\varrho(t)$. But \pref{LindbladCartoon} also implies that $\varrho(t)$ and $\varrho(t')$ only differ by terms further suppressed by $\scrH_{\rm int}$ (in the interaction picture) and so can be neglected at leading order when evaluating $\scrL_2$. This is actually a correct argument provided one never takes $t$ too different from $t'$. But it is suspicious at late times, when $t$ and $t'$ are {\it very} different because this is precisely where perturbation theory is generically unreliable (and so needs resumming). No matter how small $\scrH_{\rm int}$ is compared with $\scrH_0$ there is eventually a time for which $e^{-i \scrH_{\rm int} t}$ is not well-approximated by $1 -i \scrH_{\rm int} t$. The justification for resummation {\it at late times} must rely on some other reasoning besides perturbation theory itself. 

In what follows we argue that the Lindblad approximation during inflation is controlled for super-Hubble modes by powers of $k/(aH) \ll 1$ alone and so evolution can be Markovian, at least for super-Hubble modes. The Markovian limit emerges for super-Hubble modes because for them the system fields evolve sufficiently slowly relative to the correlation time of the environment (basically the Hubble scale). This hierarchy of scales is what underlies the relative simplicity of the Markovian limit and explains why this is a good approximation for the full evolution even at late times. Open EFT methods are useful in this context because they were introduced precisely to exploit this hierarchy (for details see the review \cite{Burgess:2022rdo}).

\subsubsection{The important interactions}

The detailed form of all cubic interactions for inflaton-metric fluctuations is given for inflationary models in near-de Sitter geometries in \cite{Maldacena:2002vr} (and summarized in eq.~\pref{allscalarcubics} of Appendix \ref{App:operators}). Concentrating on scalar perturbations\footnote{Ref.~\cite{Burgess:2022nwu} also has partial results for tensor modes, computing how tensor modes decohere scalar modes and vice versa. We here postpone a discussion of tensors to an up-coming comprehensive treatment of tensor modes.} all but three of these are suppressed for super-Hubble modes by additional factors of slow-roll parameters. In terms of the curvature perturbation $\zeta(\bm{x},t)$ -- defined more precisely in \pref{metricsplit} -- these have the schematic form
\begin{equation} \label{4ints}
   \zeta\, (\partial \zeta )^2 \,, \qquad \zeta \, \dot \zeta^2  \qquad \hbox{and} \qquad \dot \zeta \, \partial^i \zeta  \, \partial_i \chi  \,,
\end{equation}
where $\chi$ satisfies $\partial^2 \chi  = \dot \zeta$. Here over-dots denote derivatives with respect to cosmic time and $\partial_i$ contains only spatial derivatives. 

Our interest is when the environment (by assumption) involves shorter wavelength modes than system modes, so we write $\zeta = \zeta_{(s)} + \zeta_{(e)}$ (and similarly for the tensor fluctuation) where the momenta contributing to the environmental component $\zeta_{(e)}$ are systematically much larger than those appearing in the system component $\zeta_{(s)}$. But because momentum conservation requires the sum of momenta in any vertex to vanish there are no cubic interactions of schematic form $\zeta_{(s)}^2 \zeta_{(e)}$, involving two system fields and only one environment field (a triangle cannot be built with one long side and two short ones). The only cubic interactions that couple system to environment therefore have the schematic form $\zeta_{(s)} \zeta_{(e)}^2$, with derivatives acting on all possible pairs of fields.

Each of the interactions of \pref{4ints} generates a number of terms of this type. For instance the first interaction gives two types of terms, with the schematic form
\bea \label{vlist1}
   \zeta\, \partial^i \zeta \, \partial_i \zeta &=& [\zeta_\sys + \zeta_\env] \partial^i [ \zeta_\sys + \zeta_\env] \partial_i [\zeta_\sys + \zeta_\env] \nn\\
   &=& \zeta_{(s)}\, \partial^i \zeta_{(e)} \, \partial_i \zeta_{(e)}  +2 \zeta_{(e)} \, \partial^i \zeta_{(e)} \, \partial_i \zeta_{(s)}  +\cdots \,,
\eea
where the ellipses contain all other terms (that are irrelevant to decoherence at leading order). An identical argument shows
\be \label{vlist2}
   \zeta\, \dot\zeta^2 =  \zeta_\sys\, \dot\zeta_\env^2  +2 \zeta_\env \, \dot \zeta_\env \, \dot \zeta_\sys  +\cdots \,.
\ee

In \cite{Burgess:2022nwu} only the first of these terms -- the ones with derivatives acting only on the environmental component -- were kept, with the motivation that the derivatives are larger when they act on the shorter-wavelength environment fields. We do not make this restriction here, keeping instead all combinations of derivatives.  

All of the leading cubic interactions among scalar modes have coefficients $\sqrt{\slrl}/\Mp$ once expressed in terms of the canonically normalized Mukhanov-Sasaki field $v = a \Mp \sqrt{2\slrl}\, \zeta$ and so the decoherence rate found by integrating this equation is (in order of magnitude) of order $\slrl (H/\Mp)^2$. For simple inflationary models this puts an upper bound on the overall coefficient appearing in the decoherence rate:
\begin{equation}\label{prefactor}
\frac{\slrl  H^2}{8\pi \Mp^2} \sim \slrl ^2
\, \mathcal{P}_\zeta \; \lesssim 10^{-4} \times 10^{-10}
\end{equation}
where $\mathcal{P}_\zeta(k)\simeq H^2/(8\pi^2 \slrl \Mp^2) \sim 10^{-10}$ is the observed size of scalar perturbations and $\slrl \lsim 10^{-2}$ is bounded above by the non-observance of primordial tensor perturbations \cite{Planck:2018jri}. 

But this small coefficient is abundantly compensated in the full expression \pref{LindbladPurity5} by the additional factor $(aH/k)^4$ that is large in the super-Hubble limit $k \ll aH$. Furthermore this additional factor grows like $e^{4Ht}$ during inflation since $a \propto e^{Ht}$. This growth is stronger than the $a^{3Ht}$ behaviour found in \cite{Burgess:2022nwu} because the dominant contribution comes from the nonlocal interactions not considered in that paper.\footnote{Appendix \ref{App:Comparison} provides a detailed comparison between our calculation and that of \cite{Burgess:2022nwu}, highlighting why it is harder to properly identifying the leading late-time Markovian behaviour if you use the `wrong' variables. Note added: the leading power of $z$ found here (and its coefficient) differs from what was found in version 1 of this paper, and this difference can be traced to a subtlety in the ordering of limits taken when performing the integrations in the environmental correlation functions (see \cite{ToAppear} for details).} The growth is so strong that the corrections to the purity are no longer small after a handful of $e$-foldings in the super-Hubble regime, indicating a breakdown of perturbative methods. 

Beyond this point a resummation is required in order to track how the purity evolves, and this can be done along the lines discussed in \cite{Burgess:2015ajz, Burgess:2022nwu, Burgess:2024eng} by taking advantage of the Markovian nature of the leading order evolution in the super-Hubble regime. Once this is dones the result can easily overwhelm even the very small prefactor given in \pref{prefactor} over the 50 $e$-foldings of inflation available in inflationary models. We set up the tools required to perform this resummation and discuss the complications that led us to defer it to future work.

Our presentation is structured as follows. To start off \S\ref{sec:OpenEFT} reviews the fluctuations within inflationary models and their cubic interactions that relevant for the later decoherence calculations. \S\ref{sec:SysEnv} then sets up the description of system and environment for these modes, with the system chosen to include the range of wave numbers observable in cosmology and the environment describing those with much shorter wavelengths. All of these discussions are review material that can be skipped by the cogniscenti. 

The guts of the calculation begin in \S\ref{sec:markstrat} where the time-dependence of the important correlation functions is isolated. This is done in a formally exact way, expanding in the correlation time but without truncating the sum. This allows a systematic assessment of the size of any particular term in powers of the small quantities in the calculation. \S\ref{sec:SHL} then sketches the calculation of the various correlators that arise in the evolution of scalar-metric modes, leading to expressions like \pref{FinalSmallzzetazeta} for how they depend on the super-Hubble expansion parameter $z = - k\eta = k/(aH)$. This allows the determination of the dominant evolution for super-Hubble correlations, including an assessment of what controls their Markovianity. \S\ref{sec:PurityCalc} then applies these expressions to compute the evolution of the purity, culminating in the final result \pref{LindbladPurity5} for the perturbative purity evolution. We summarize the implications of our calculation in \S\ref{sec:Conclusions} with a brief discussion of the open ends that our calculation does not resolve and possible next steps. Several useful side arguments are grouped into appendices.

\section{Open system of scalar metric modes}
\label{sec:OpenEFT}

This section briefly summarizes our calculational setup (we refer the reader to \cite{Burgess:2022nwu} for more details). 

The system of interest is the simplest `single-clock' inflationary models, with the metric $g_{\mu\nu}$ coupled to a real scalar inflaton field $\varphi$ through the action\footnote{We use MTW conventions \cite{MTW} in this paper.}
\begin{equation}
\label{actionstart}
S = \int \exd^4 x\; \sqrt{ - g} \bigg[\tfrac12  \Mp^2 \, R
    - \tfrac{1}{2} g^{\mu\nu} \, \partial_{\mu} \varphi \,
    \partial_{\nu} \varphi  - V(\varphi) \bigg]
\end{equation}
where $R$ is the Ricci scalar and $V(\varphi)$ is the potential energy of the inflaton $\varphi$.  Our focus is on the late-time evolution of fluctuations about a homogeneous spatially flat FRW geometry
\begin{equation}
\label{metric}
\varphi = \phi(t) \qquad \hbox{with metric} \qquad
\exd  s^2 = - \exd  t^2 + a^2(t) \, \exd  \bm{x}^2 = a^2(\eta)
  \left( - \exd  \eta^2 + \exd \bm{x}^2 \right) \,.
\end{equation}
As usual, cosmic time ($t$) and conformal time ($\eta$) are related by $\exd t = a \, \exd \eta$, and throughout the paper we use overdots to denote differentiation with respect to $t$ and primes to denote differentiation with respect to $\eta$.

In the strict de Sitter limit the scale factor is given by
\begin{equation}
\label{eq:defscalefactor}
a \simeq e^{H_\ssI t} \simeq-\frac{1}{H_\ssI \eta} \,,
\end{equation}
for constant $H_\ssI$, with $-\infty < t < \infty$ corresponding to $- \infty < \eta < 0$. In practice we only ask the evolution to be close to de Sitter, in the sense that the scale factor $a$ and background scalar $\phi$ evolve slowly so the Hubble scale $H = \dot a/a$ has a small time derivative: $\slrl =-{\dot{H}}/{H^2} \ll 1$ and so $H(t) \simeq H_\ssI$ is approximately constant. The $\phi$ field equation implies the scalar field's background value in this approximation satisfies 
\be\label{phidotvseps}
\dot{\phi}^2=2H^2\Mp^2\slrl , 
\ee
and so its kinetic energy is also small compared to its potential energy.  

Fluctuations about this background are described by expanding the scalar field and metric
\begin{eqnarray} \label{ADMmetric}
\varphi = \phi(t) + \delta \varphi(t,\bm{x}) \quad\hbox{and} \quad
  \exd  s^2 = - N^2 \exd  t^2 + h_{ij} \big( \exd  x^i + N^i \exd  t \big)
  \big( \exd  x^j + N^j \exd  t \big) \,,
\end{eqnarray}
and picking a gauge to fix time and spatial reparametrizations. Standard arguments show that using this expansion in the action (\ref{actionstart}) ends up leaving a single physical scalar degree of freedom plus the two tensor modes describing gravitational waves. Following \cite{Maldacena:2002vr} we write
\begin{equation} \label{metricsplit}
     h_{ij} = a^2 e^{2\zeta} \hat h_{ij} \quad\hbox{with}\quad
     \hat h_{ij} = \delta_{ij} + \gamma_{ij} + \tfrac12 \,
     \delta^{kl} \gamma_{ik} \gamma_{lj} + \cdots \,,
\end{equation}
where $\det \hat h_{ij} = 1$ and $\delta^{ij} \partial_i \gamma_{jk} = \delta^{ij} \gamma_{ij} = 0$. The lapse $N$ and shift $N^i$ are determined by solving the energy and momentum constraints and the remaining two scalar functions $\delta \varphi$ and $\zeta$ are not independent, as can be seen by using a coordinate transformation to switch between two convenient gauge choices: $\delta \varphi = 0$ (co-moving gauge) or $\zeta = 0$ (spatially-flat gauge). 

\subsection{Scalar quadratic action}

Working in the co-moving gauge and temporarily dropping tensor fluctuations, the part of the action governing scalar modes can be expanded in powers of $\zeta$, with $S = {}^{(2)}S + {}^{(3)}S + {}^{(4)}S + \cdots$ where ${}^{(n)}S$ involves $n$ powers of the fluctuation fields. The leading (quadratic) part of the action governing scalar fluctuations has the form (see for example \cite{Kodama:1985bj, Mukhanov:1990me, Maldacena:2002vr})
\be 
\label{freescalaraction}
  ^{(2)}S = \int \exd  t \; \exd^3 \bm{x}\;
  \frac{\dot{\phi}^2}{2H^2}\bigg[ a^3 \dot{\zeta}^2
    - a (\partial \zeta )^2 \bigg]   = \int \exd  \eta \; \exd^3 \bm{x}\;
  {\slrl} \Mp^2 a^2\bigg[ (\zeta')^2 -  (\partial \zeta )^2 \bigg]   \,,
\ee
where the second equality trades $\dot\phi^2/H^2$ for the slow-roll parameter using \pref{phidotvseps} and changes variables to conformal time, using $\zeta'  = a\dot\zeta$. 
As usual $(\partial \zeta)^2 = \delta^{ij} \partial_i \zeta \, \partial_j \zeta$.

Following standard practice, we work in semiclassical perturbation theory,  
\begin{equation}
\label{eq:Hamiltonian:cubic}
\scrH(\eta) = \scrH_0(\eta) + \scrH_{\mathrm{int}}(\eta)
\end{equation}
where $\scrH_0$ is constructed from the quadratic part of the action ${}^{(2)}S$ and the interaction $\scrH_{\rm int}$ built using ${}^{(n)}S$ for $n \geq 3$. We use the interaction picture, for which the fields satisfy their free equations of motion. For evolution in conformal time the momentum conjugate to $\zeta$ as obtained from the free Hamiltonian is 
\be \label{pzeta0}
   \mfp = \frac{\delta S }{ \delta \zeta'} = 2{\slrl}\Mp^2 a^2\zeta' \,, 
\ee
and so the quadratic Hamiltonian in canonical form is
\begin{equation}
\label{freeHclassicalx}
\scrH_0 := \int\exd^3 \bm{x} \; p \, \zeta' - \scrL_0 =  \int\exd^3 \bm{x} \; \left[\frac{ \mfp^2}{4\slrl \Mp^2 a^2} + \slrl \Mp^2 a^2 (\partial \zeta)^2 \right]\,.
\end{equation}

\subsubsection{Mukhanov-Sasaki variables}

To diagonalize this Hamiltonian it is useful first to remove the time-dependent coefficients by switching to the canonical Mukhanov-Sasaki variable~\cite{Mukhanov:1981xt,Kodama:1985bj} 
\begin{equation}
\label{eq:defzeta}
v(\eta, {\bm x}) := \mfz \, \zeta(\eta,{\bm x}) \qquad \hbox{where} \quad \mfz := a\Mp \sqrt{2\slrl } \,,
\end{equation}
in terms of which the quadratic action \pref{freescalaraction} becomes (see for example \cite{Kodama:1985bj, Mukhanov:1990me, Maldacena:2002vr})
\be
\label{freescalaractionv}
  ^{(2)}S   = \tfrac{1}{2}
\int{\exd  \eta\; \exd^3 \bm{x} \;
\bigg[\left(Dv\right)^2
- (\partial v )^2
  \bigg]}  
 = \tfrac{1}{2}
\int{\exd  \eta\; \exd^3 \bm{x} \;
\bigg[\left(v^\prime\right)^2
-(\partial v )^2
+\frac{\mfz''}{\mfz } v^2 \bigg]} + \hbox{s.t.}\,,
\ee
where 
\begin{equation}
 Dv := v' - \frac{\mfz' v}{\mfz} 
\end{equation}
and the second equality performs an integration by parts, for which the corresponding surface terms are denoted by `s.t.'.
 
It is worth commenting on the integration by parts since we later ask whether calculations of decoherence are equivalent when done using either of the actions shown in eqs.~\pref{freescalaraction} and \pref{freescalaractionv}. Integration by parts in the time direction leads to `boundary' or surface contributions in the action evaluated at the initial and final times, $t_f$ and $t_i$:
\be
 \hbox{s.t.} = \int_{t_i}^{t_f} \exd t  \; \partial_t F(t) =   F(t_f) - F(t_i) \qquad \hbox{where} \qquad F =- \tfrac12 \int  \exd^3 \bm{x} \left( \frac{\mfz'}{\mfz } v^2 \right)\,.
\ee
These surface terms can be interpreted as contributions to the vacuum wave-functional in a path-integral amplitude:\footnote{An identical argument applies to an `in-in' amplitude once one keeps track of the two time contours.}
\be
  \int \cD \phi_i \, \cD\phi_f \Psi^\star[\phi_f] \, \langle \phi_f | \phi_i \rangle \, \Psi_0[\phi_i] 
  =  \int \cD \phi_i \, \cD\phi_f \int_{\phi_i}^{\phi_f} \cD \phi \, e^{i S} \, \Psi^\star[\phi_f]  \, \Psi_0[\phi_i] 
\ee
with $\scrL \to \scrL + \partial_t F$ being equivalent to $\Psi[\phi] \to e^{-i F[\phi]}\Psi[\phi]$ at both initial and final times. Using the action after the integration by parts makes the system into a harmonic oscillator provided one keeps in mind that the states have also been modified -- or squeezed -- by the transformation $e^{-iF}$. 

This same conclusion can also be drawn without integrating by parts, directly using the action given in the first equality of \pref{freescalaractionv}. To see how, define the canonical momentum and Hamiltonian by
\be \label{Ddef}
   \tilde p := \frac{\delta S}{\delta v'} = Dv = v' - \frac{\mfz'}{\mfz} \, v \,,
\ee
and so
\be \label{squeezedfree}
 \widetilde \scrH_0 = \int\exd^3 \bm{x} \;  \tilde p v' - \scrL_0 =   \int\exd^3 \bm{x} \; \left[\tfrac12 \tilde p^2 + \tfrac12 (\partial v)^2 + \frac{\mfz'}{\mfz} \, \tilde p v\right]\,.
\ee  
The $\tilde p\,$-$v$ term can be removed by performing the canonical transformation $\tilde p \to p = \tilde p + (\mfz'/\mfz) v$ and $v \to v$. The commutator $[\tilde p(\bm{x}), v(\bm{x}')] = -i \delta^3(\bm{x} - \bm{x}')$ shows this is accomplished by the transformation $v \to U^\dagger v \,U$ and $\tilde p \to U^\dagger \tilde p \, U$ with
\be
   U = \exp\left[ \tfrac{i}{2} \int \exd^3 \bm{x} \left( \frac{\mfz'}{\mfz} \, v^2 \right) \right] = e^{-iF} \,,
\ee
showing that the states are related in the same way as found when integrating by parts. Applying this transformation to the time-dependent Schr\"odinger equation -- time-dependent because the transformation $U$ is time-dependent -- shows that the transformed state evolves with the free Hamiltonian 
\begin{equation}
\label{freeHclassicalx}
\scrH_0 := \tfrac{1}{2} \int\exd^3 \bm{x} \; \left[ p^2+
( \partial v  )^2 - \frac{\mfz^{\prime\prime}}{\mfz} \; v^2 \right] \qquad \hbox{with} \qquad p = v' \,,
\end{equation}
precisely as would have been found from the integrated-by-parts action from \pref{freescalaractionv}. (See \cite{Braglia:2024zsl} for a detailed discussion of surface terms and canonical transformations in cosmology, including for the interaction terms we use below.)

All roads lead to Rome and this free Hamiltonian $\scrH_0$ is diagonalized by going to momentum space (box normalized in a comoving volume $V$),
\begin{equation}
\label{eq:v:Fourier}
v(\eta,\bm{x}) =\frac{1}{\sqrt{V}}\sum_{\bm{k}}\; 
v_{\bm{k}}(\eta) \, e^{i \bm{k}\cdot\bm{x}}\,,
\end{equation}
where hermiticity in real space $v(\eta,\bm{x}) = v^\dagger(\eta,\bm{x})$ implies $ v_{-\bm{k}}(\eta) = v_{\bm{k}}^\dagger(\eta)$ and a similar expression holds for the conjugate momentum $ p_{\bm{k}}$. We write $ v_{\bm{k}}(\eta)$ in terms of mode functions $u_{{k}}(\eta)$
\begin{equation} \label{vhatk}
 v_{\bm{k}}(\eta) = u_{{k}}(\eta) {c}_{\bm{k}}
  + u^{\ast}_{{k}}(\eta) {c}^{\dagger}_{-\bm{k}} \,.
\end{equation}
where we assume rotation invariance so $u_k(\eta)$ depends only on $k = |\bm{k}|$. Normalizing the modes using $u_{{k}} u_{{k}}^{*\prime}-u_{{k}}^* u_{{k}}'=i$ ensures the equal-time commutation relations 
\begin{equation}
\label{equaltimeCCR}
{\big[} { v}_{\bm{k}}(\eta),  p_{\bm{q}}(\eta) \big] = i \delta_{\bm{k},- \bm{q}},
\end{equation}
are equivalent to $[{c}_{\bm{k}}, {c}^{\dagger}_{\bm{q}}] = \delta_{\bm{k},\bm{q}}$. 

With these choices 
\begin{equation}
\label{freeHclassical}
\scrH_0(\eta)  :=  \frac{1}{2}\sum_{\bm{k}} \;
\Bigl[ p_{\bm{k}}(\eta)  p_{-\bm{k}}(\eta)
+ \omega^2(\bm{k},\eta)  v_{\bm{k}}(\eta)  v_{-\bm{k}} (\eta) \Bigr] \,,
\end{equation}
where the time-dependent frequency is
\begin{equation}
\label{omegadef}
\omega^2(\bm{k},\eta) := k^2 - \frac{\mfz^{\prime\prime}}{\mfz} \ .
\end{equation}
In the limit $\slrl \to 0$ the scale factor $a$ is given by \pref{eq:defscalefactor}  for $- \infty < \eta < 0$, leading to the well-known de Sitter form $\omega^2(\bm{k},\eta) \simeq k^2 - ({2}/{\eta^2})$.

\subsection{Scalar cubic self-interactions}
 
The cubic scalar self-interactions predicted by the action \pref{actionstart} that arise at leading order in $1/\Mp$ are \cite{Maldacena:2002vr} (see Appendix \ref{ssec:AppCubic})
\be
\label{cubiczeta}
{}^{(3)}S_{\rm lo} 
  =  \int \exd  \eta\, \exd^3 \bm{x}\;\slrl ^2 \Mp^2 a^2 \Biggl\lbrace
   \left(\partial \zeta\right)^2  \zeta
+  ({\zeta'})^2 \zeta  - 2  
{\zeta'} \left(\partial_i \partial^{-2} {\zeta'}\right)
\left(\partial_i \zeta\right) \Biggr\rbrace\,, 
\ee 
with all other cubic interactions -- {\it c.f.}~eq.~\pref{allscalarcubics} -- being more suppressed in the slow-roll expansion. We require the interaction Hamiltonian that follows from this expression.

The terms involving $\zeta'$ modify the expression for the canonical momentum from the leading result $\mfp_0 = 2{\slrl}\Mp^2 a^2\zeta'$ given in \pref{pzeta0} to
\be
     \mfp 
   = 2\slrl \Mp^2 a^2 \Biggl\lbrace \zeta' + \slrl \Bigl[\zeta'  \zeta  -  \left(\partial_i \partial^{-2} {\zeta'}\right) \left(\partial_i \zeta\right)
   +\partial^{-2} \partial_i  (\zeta' \partial_i \zeta) \Bigr] \Biggr\rbrace \,,
\ee
and so
\bea
   \zeta'
   &\simeq&\frac{1}{2\slrl \Mp^2 a^2} \Biggl\lbrace \mfp -  \slrl \Bigl[\mfp  \zeta  -  \left(\partial_i \partial^{-2} {\mfp}\right) \left(\partial_i \zeta\right)
   +\partial^{-2} \partial_i  (\mfp \partial_i \zeta) \Bigr] \Biggr\rbrace\\
   &=& \frac{\mfp}{2\slrl \Mp^2 a^2}+ \delta \zeta'\,, \nn
\eea
where the first line drops the difference between $\mfp_0$ and $\mfp$ in the correction term and the second line defines $\delta \zeta'$.

The total Hamiltonian $\scrH = \int \exd^3 \bm{x} \, \mfp \zeta' - \scrL$ is then constructed as usual and the interaction Hamiltonian obtained by subtracting $\scrH_0$ -- as given in \pref{freeHclassicalx} -- from the result:
\bea
\label{Hclassicalint}
\scrH &=& \int \exd^3 \bm{x} \;  \mfp   \zeta' - \scrL \nn\\
&=&  \int \exd^3 \bm{x} \Biggl\lbrace \mfp \left[ \frac{\mfp}{2{\slrl}\Mp^2 a^2} + \delta \zeta' \right] -  \slrl \Mp^2 a^2 \left[ \frac{\mfp}{2{\slrl}\Mp^2 a^2} + \delta \zeta' \right]^2 +   \slrl \Mp^2 a^2 (\partial \zeta)^2 \nn\\
&& \qquad\qquad\qquad\qquad  - \slrl ^2 \Mp^2 a^2  
  \left(\partial \zeta\right)^2  \zeta
 - \frac{1}{4 \Mp^2 a^2} \Bigl[  \mfp^2 \zeta  - 2  
\mfp \left(\partial_i \partial^{-2} \mfp \right)
\left(\partial_i \zeta\right) \Bigr]  \Biggr \rbrace \\
&=& \scrH_0 +\scrH_{\rm int} \,, \nn
\eea
where
\be\label{Hintzeta}
  \scrH_{\rm int} = -  \int \exd^3 \bm{x} \Biggl\lbrace    \slrl ^2 \Mp^2 a^2  
  \left(\partial \zeta\right)^2  \zeta +  \frac{1}{4 \Mp^2 a^2} \Bigl[  \mfp^2 \zeta  - 2  
\mfp \left(\partial_i \partial^{-2} \mfp \right)
\left(\partial_i \zeta\right) \Bigr]   \Biggr \rbrace \ .
\ee
 
The counting of powers of $\slrl$ and $1/\Mp$ in perturbative results is more easily seen when the interactions are expressed in terms of the variable $v$ and its canonical momentum $p$ because their correlation functions are $\slrl$- and $\Mp$-independent. One finds in this way the interaction lagrangian
\be 
\label{cubiczetav}
{}^{(3)}S_{\rm lo}  \simeq  \frac{1}{2\sqrt2 \, \Mp}
  \int \exd  \eta\, \exd^3 \bm{x}\, \frac{\sqrt{\slrl}}{a} \Biggl\lbrace  \left[ v \left(\partial v\right)^2   
+ v (D v)^2  \right] - 2   
Dv \left(\partial_i \partial^{-2}Dv\right)
 \partial_i v  \Biggr\rbrace \,, 
\ee
showing that all terms are suppressed by $\sqrt{\slrl}/\Mp$ relative to the leading Hamiltonian given for these variables in \pref{squeezedfree} or \pref{freeHclassicalx}. Repeating the steps leading to the interaction Hamiltonian then gives
\be 
\label{cubiczetavH}
 \scrH_{\rm int}  \simeq - \frac{1}{2\sqrt2 \, \Mp}
  \int \exd  \eta\, \exd^3 \bm{x}\, \frac{\sqrt{\slrl}}{a} \Biggl\lbrace  \left[ v \left(\partial v\right)^2   
+ v \tilde p^2  \right] - 2   
\tilde p \left(\partial_i \partial^{-2} \tilde p\right)
 \partial_i v  \Biggr\rbrace \,.
\ee

Although the interaction Hamiltonian as given above is obtained in co-moving gauge (and gauge-independent observables are difficult to construct at cubic and higher order in cosmological perturbation theory), \cite{Maldacena:2002vr} checked that a calculation performed in the spatially-flat gauge gives the same result for the bispectrum. We take this as evidence that physical results obtained from our calculations using this interaction Hamiltonian will be gauge
independent.

\section{The system and the environment}
\label{sec:SysEnv}

\begin{figure}
\centering
  \includegraphics[width=0.65\linewidth]{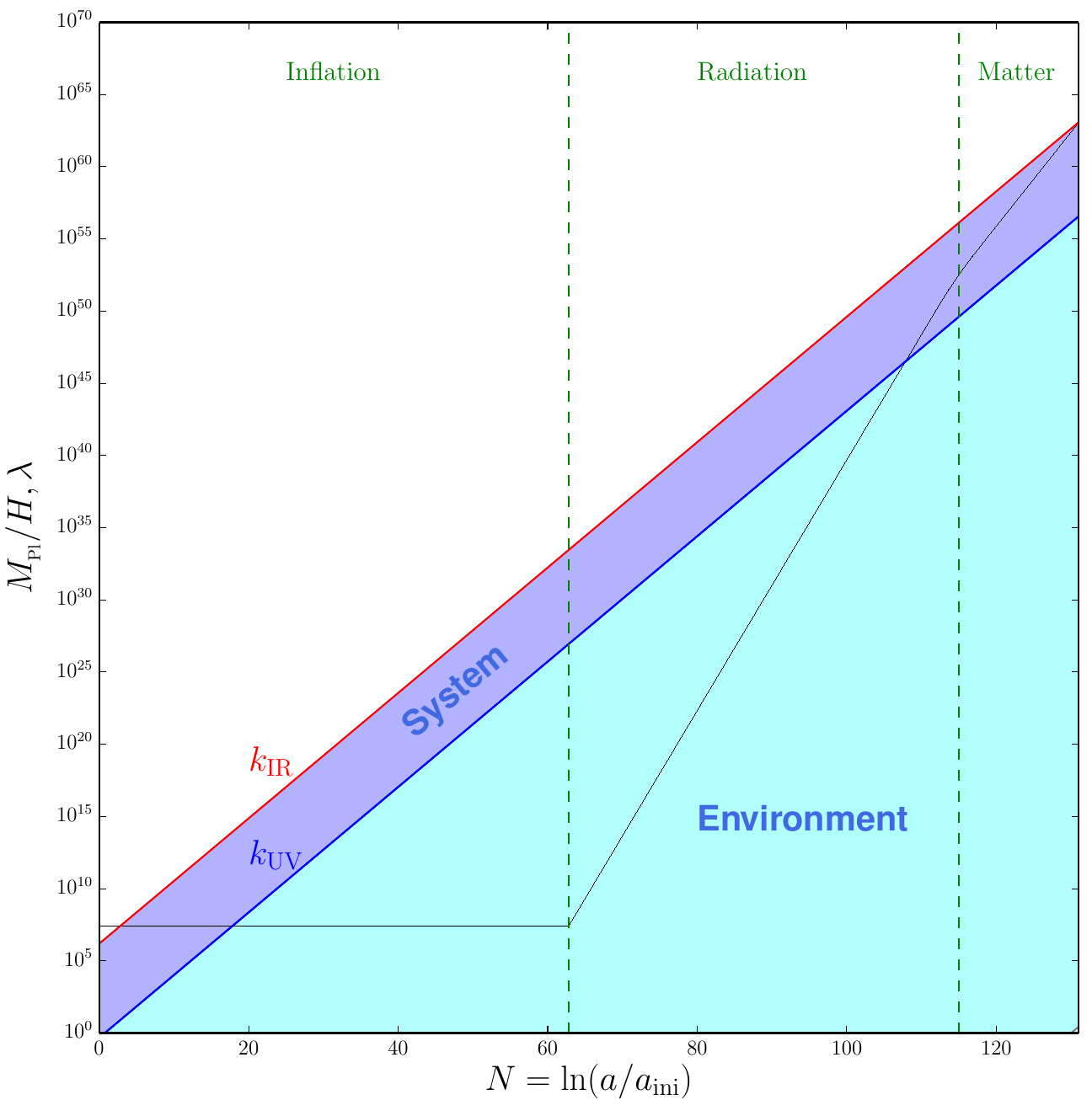}
  \caption{\small A sketch of the domain of the system and environment modes described in the text. The black line denotes the Hubble radius and the coloured
    lines stand for the mode wavelengths. The system consists of co-moving scales between $k_{\IR}$ and $k_{\UV}$, both of which are outside the Hubble radius at the end of inflation. The environment is made of all scales such that $k>k_\UV$. (Figure taken from \cite{Burgess:2022nwu}.)
   }
\label{fig:defsystem}
\end{figure}

We follow \cite{Burgess:2022nwu} and divide the space of field states into a `system' consisting of those modes that appear in observations made at late times and an `environment' consisting of all shorter-wavelength modes that are not observed. We do so using comoving momenta, so the boundary $k_\UV$ separating system from environment does not evolve in time, though the corresponding physical scale $k_\UV/a$ does (see Fig.~\ref{fig:defsystem}). Having made this split we trace out over the environment modes and ask how they affect state evolution within the observed sector. In practice, present-day measurements only sample primordial fluctuations with co-moving momenta lying within a range 
\begin{equation}
\label{eq:defsystemintermsofk}
k_{\IR} < k < k_{\UV} 
\end{equation} 
where $k_{\IR}/a_0 \sim 0.05 \, a_0 \mathrm{Mpc}^{-1}$ and $k_{\UV}/a_0$ (with $k_\UV \sim 2500 \, k_\IR$) are the smallest and
largest currently observable physical momenta (such as through CMB or large-scale structure observations) and $a_{0}$ is the
present-day scale factor. 

In practice (for simplicity) we lump all modes with $k < k_\IR$ whose wavelengths are too long to have been observed in with the system, though this does not affect the decoherence rate we ultimately compute. Denoting the observed system by $(s)$ and the environment by $(e)$, we write the position-space field as
\begin{equation}
\label{vdecomp}
      {\zeta}(\eta,\bm{x}) = {\zeta}_{\sys}(\eta,\bm{x}) \otimes {\mathcal{I}}_{\env} 
      +  \mathcal{I}_{\sys} \otimes {\zeta}_{\env}(\eta,\bm{x})
\end{equation}
where $\cI_\sys$ and $\cI_\env$ represent the appropriate unit operators. The system and environmental fields appearing here are defined by
\begin{equation}
\label{vclassicalsplit}
\zeta_{\sys}(\eta,\bm{x})  : = 
 \frac{1}{\sqrt{V}}\sum_{k<k_{UV}}\; \;
\zeta_{\bm{k}}(\eta) e^{i \bm{k}\cdot\bm{x}} \quad \hbox{and} \quad
\zeta_{\env}(\eta,\bm{x})  : = 
  \frac{1}{\sqrt{V}}\sum_{k>k_{UV}}\; \;
\zeta_{\bm{k}}(\eta) e^{i \bm{k}\cdot\bm{x}} \,.
\end{equation}
Under this decomposition the free Hamiltonian becomes
\begin{equation}
  {\scrH}_0(\eta) = {\scrH}_{\sys}(\eta) \otimes {\mathcal{I}}_{\env}
  + {\mathcal{I}}_{\sys} \otimes {\scrH}_{\env}(\eta)
\end{equation}
where $\scrH_\sys$ and $\scrH_\env$ are both given by
\pref{freeHclassical} but with the momentum range respectively
restricted to the intervals $k < k_\UV$ and $k > k_\UV$.

Decoherence is driven by those terms in the Hamiltonian that connect the system and environment and these are obtained by inserting the decomposition (\ref{vdecomp}) into the cubic interaction Hamiltonian $\scrH_{\mathrm{int}}$ defined in \pref{Hclassicalint}. This gives contributions of the schematic form $\zeta_{\sys}^3$, $\zeta_{\sys}^2 \zeta_{\env}$, $\zeta_{\sys} \zeta_{\env}^2$ and $\zeta_{\env}^3$. As argued in \S\ref{sec:intro}, only the cross terms ($\zeta_{\sys}^2 \zeta_{\env}$ and $\zeta_{\sys} \zeta_{\env}^2$) couple the system to the environment and of these momentum conservation suppresses the $\zeta_{\sys}^2 \zeta_{\env}$ interactions because it is impossible to sum two small momenta to get a large one. We keep all terms with derivatives acting on both system and environment fields, unlike \cite{Burgess:2022nwu} which focussed on terms for which derivatives act only on environmental modes. Applied to the Hamiltonian \pref{Hintzeta} this leads to the following interaction Hamiltonian
\bea\label{Hintmom}
  \scrH_{\rm int} &=& -  \int \exd^3 \bm{x} \Biggl\lbrace    \slrl^2 \Mp^2 a^2  
  \left[\partial \zeta_\env\right]^2  \zeta_\sys +  2 \slrl ^2 \Mp^2 a^2  
  \partial_i \zeta_\env \partial_i \zeta_\sys  \zeta_\env  \nn\\
  && \qquad \qquad +  \frac{1}{4 \Mp^2 a^2} \Bigl[  \mfp_\env^2 \zeta_\sys +  2\mfp_\env \mfp_\sys \zeta_\env   - 2  
\mfp_\env \left(\partial_i \partial^{-2} \mfp_\env \right)
\left(\partial_i \zeta_\sys\right)  \\
&& \qquad\qquad\qquad - 2  
\mfp_\env \left(\partial_i \partial^{-2} \mfp_\sys \right)
\left(\partial_i \zeta_\env\right)  - 2  
\mfp_\sys \left(\partial_i \partial^{-2} \mfp_\env \right)
\left(\partial_i \zeta_\env \right) \Bigr]   \Biggr \rbrace \,. \nn
\eea
For the interaction picture the fields $\zeta_\sys(\bm{x},\eta)$ and $\zeta_\env(\bm{x},\eta)$ are both evaluated as functions of time using the free evolution equations. 

\subsection{System evolution}
\label{sec:conversion}

The next step is to predict the evolution of the system's reduced density matrix, ${\varrho}(\eta)$, obtained by tracing out the environmental degrees of freedom: 
\begin{equation} \label{reducedSpic}
{\varrho}(\eta) := \Trenv \Bigl[ {\rho}(\eta) \Bigr] \,,
\end{equation}
where $\rho$ is the density matrix for the full system. 

For inflationary applications the full system's state is usually chosen initially to be in a pure state, usually chosen to be the Bunch Davies vacuum \cite{Bunch:1978yq}, $|\Omega \rangle = |0 \rangle_\sys \otimes |0\rangle_\env$, so
\begin{equation}
\label{ICfull}
{\rho}(\eta_{\mathrm{in}}) 
= | \Omega \rangle \langle \Omega | = | 0_{\sys} \rangle \langle 0_{\sys} |
\otimes  | 0_{\env} \rangle \langle 0_{\env} | \,,
\end{equation}
where
\begin{equation}
\label{vacuumsplitting}
| 0_{\sys} \rangle := \bigotimes_{k<k_{\UV} } | 0_{\bm{k}} \rangle
\quad  \mathrm{and} \quad | 0_{\env} \rangle :=
\bigotimes_{ k>k_{\UV} } | 0_{\bm{k}} \rangle \quad \hbox{with} \quad
{c}_{\bm{k}}(\eta_{\mathrm{in}}) | 0_{\bm{k}} \rangle = 0
\quad \mathrm{for\ all\ }\bm{k} \,.
\end{equation}
With this choice the mode functions $u_{{k}}(\eta)$ appearing in \pref{vhatk} are (for massless states on a de Sitter background)
\be \label{modev}
  u_k(\eta) = \frac{1}{\sqrt{2k}} \left( 1 - \frac{i}{k\eta} \right) e^{-i k \eta} \,,
\ee
and so the corresponding modes appearing in $\zeta_\bmk$ are given by 
\begin{equation}
\label{eq:deSitter:BD:vk}
\hat u_{{k}}(\eta) = \frac{u_k(\eta)}{\mfz(\eta)} = \frac{i }{\sqrt{4 k^3 \slrl}} \left( \frac{H}{\Mp}\right)\left(1+i k\eta\right)e^{-i k\eta}\, ,
\end{equation}
and the initial reduced density matrix factorizes
\begin{equation} \label{rhofactor}
{\varrho}(\eta_{\rm in}) = \bigotimes_{k < k_{\UV} }
{\varrho}_{\bm{k}}(\eta_{\rm in})  \,.
\end{equation}
This is a natural choice for the initial state provided it is made in the remote past -- {\it i.e.}~in the limit $\eta_{\rm in} \to - \infty$.

Subsequent state evolution in the full theory is given in the Schr\"odinger picture by the Liouville equation 
\begin{equation} \label{SpicVN}
\frac{\partial {\rho}_{\ssS}}{\partial \eta} = - i \Bigl[ {\scrH}_{\ssS}(\eta)   , {\rho}_{\ssS}(\eta) \Bigr] \,,
\end{equation}
and so the interaction picture density matrix satisfies (see \cite{Burgess:2022nwu} for details)
\begin{equation} \label{INTpicVN}
  \frac{\partial {\rho}}{\partial \eta}
  = - i \Bigl[ {\scrH}_{\mathrm{int}}(\eta) , {\rho}(\eta) \Bigr] \,,
\end{equation}
where in practice $\scrH_{\rm int}$ is built from the interactions listed in \pref{Hintmom} (or its equivalent built from $v$ and $p$).

\subsubsection{Integrating out the environment}

The evolution of the reduced density matrix $\varrho$ is in principle obtained by taking the trace of \pref{INTpicVN}. This is not so useful at face value because the trace of the right-hand side of the equation involves both $\varrho$ and the state of the environmental degrees of freedom. A more useful expression is obtained by solving for the evolution of the environmental state and then substituting this back into the evolution equation for $\varrho$, since this makes direct reference only to the system state. But it does so at a price of introducing nonlocality in time since $\partial_t \varrho$ depends on the entire past history $\varrho(t')$ for $t' < t$. The linearity of the Liouville equation allows this integration process to be done very generally perturbatively in $\scrH_{\rm int}$, and results in what is called the Nakajima-Zwanzig equation \cite{Nakajima:1958pnl, Zwanzig:1960gvu} (for a review aimed at applications to gravity see \cite{Burgess:2022rdo}). 

The resulting evolution equation for $\varrho$ at second order in $\scrH_{\rm int}$ is derived in detail in \cite{Burgess:2022nwu} for an interaction Hamiltonian of the form
\begin{equation}
\label{HintRdef0}
      {\scrH}_{\mathrm{int}}(\eta) =  G (\eta)   \int\exd^3 \bm{x}\;
     \cS_a (\eta, \bm{x}) \otimes \cE^a(\eta,\bm{x}) \qquad \hbox{where} \qquad
     G(\eta)=-\slrl^2 \Mp^2 a^2\,,
\end{equation}
and there is an implied sum on $a$. The $\cS_a$ are a collection of operators built using $\zeta_\sys$ and its derivatives while $\cE^a$ denotes a similar collection built from $\zeta_\env$ and its derivatives.  For instance, for the interaction \pref{Hintmom} the following system operators 
\be\label{cSdefs}
   \cS_\zeta := \zeta_\sys  \,, \qquad \cS_\mfp := \mfp_\sys  \,, \qquad
   \cS_{\zeta i} := \partial_i \zeta_\sys \,, \qquad 
   \cS_{\mfp i} =   \partial_i \partial^{-2} \mfp_\sys \,,
\ee
pair off with the environmental counterparts
\be 
\label{eq:B:def1}
    \cE^\zeta    :=  \left\lbrace \left[\partial \zeta_\env\right]^2  +  \frac{\mfp_\env^2}{4  \slrl^2 \Mp^4 a^4} \right\rbrace\,,
\ee 
\be 
\label{eq:B:def2}
     \cE^\mfp   :=   \frac{2}{\mfz^4} \Bigl[  \mfp_\env \zeta_\env - 
      \left(\partial_i \partial^{-2} \mfp_\env \right)
\left(\partial_i \zeta_\env \right) \Bigr] \,,
\ee 
\be 
\label{eq:B:def3}
    \cE^{\zeta i}  :=  \delta^{ij} \left\lbrace    
  \partial_j \zeta_\env   \zeta_\env   -  \frac{\mfp_\env}{2 \slrl ^2 \Mp^4 a^4}
 \left(\partial_j \partial^{-2} \mfp_\env \right) \right\rbrace \,,
\ee 
and
\be 
\label{eq:B:def4}
     \cE^{\mfp i}   :=    - \frac{2 \delta^{ij}}{ \mfz^4} \mfp_\env \partial_j \zeta_\env   \,.  
\ee 

The Nakajima-Zwanzig evolution equation at second order in $\scrH_{\rm int}$ then is
\bea
\label{NZ}
\frac{\partial {\varrho}}{\partial \eta} &\simeq&
- i  \scrE^a(\eta) \int \exd^3\bm{x} \;
G (\eta)\ \Bigl[  \cS_a(\eta,\bm{x}),  {\varrho}(\eta) \Bigr] \nn\\
 &&\qquad -  \int \exd^3\bm{x} \int \exd^3\bm{x}' \int_{\eta_{\mathrm{in} } }^{\eta} \exd  \eta'  \; G (\eta)\; G (\eta^{\prime}) \bigg\lbrace \Bigl[  \cS_a(\eta,\bm{x}) ,
  \cS_b(\eta',\bm{x}')  {\varrho}(\eta') \Bigr] C^{ab}(\eta, \eta';
\bm{x} - \bm{x}') \nn\\
&& \qquad\qquad\qquad\qquad 
+ \Bigl[  {\varrho}(\eta') \cS_b(\eta',\bm{x}')  ,  \cS_a(\eta,\bm{x}) \Bigr]
C^{ba}(\eta', \eta;\bm{x} - \bm{x}') \bigg\rbrace 
\eea
where
\begin{equation}
\label{R_1pt_def}
\scrE^a(\eta) := \langle 0_{\env} | \cE^a(\eta,\bm{x}) | 0_{\env} \rangle
\end{equation}
and
\begin{equation}
\label{2pt_text}
C^{ab}(\eta,\eta' ; \bm{x} - \bm{x}') =  \langle 0_{\env} |
\Bigl[ \cE^a(\eta,\bm{x}) - \mathscr{E}^a(\eta) \Bigr]
\Bigl[ \cE^b(\eta',\bm{x}') - \mathscr{E}^b(\eta')  \Bigr] | 0_{\env} \rangle \,,
\end{equation}
and (when quoting the position dependence of the left-hand side) we assume an environmental state that is invariant under spatial translations and reflections (such as the Bunch-Davies state). This is also where our discussion starts to differ significantly from \cite{Burgess:2022nwu}, which up to this point would have differed by dropping all terms except $\cS_{\zeta} \otimes \cE^{\zeta}$ in \pref{HintRdef0} and omitting the $\mfp^2_\env$ term in $\cE^\zeta$ seen in \pref{eq:B:def1}. 

The time integral in \pref{NZ} can diverge in the limit $\eta' \to \eta$, a divergence whose roots lie in the short-distance singularities that correlation functions contain in the coincidence limit.  Our mission is to evaluate these correlation functions for the operators in \pref{eq:B:def1} through \pref{eq:B:def4} and use the results in \pref{NZ} to evolve the reduced density matrix and part of the story when doing so involves a discussion of how this UV divergence is handled. We shall see that their structure allows them to be renormalized in a standard way \cite{Burgess:2003jk} that does not directly enter into the leading prediction for the decoherence rate. 

A second type of divergence can also arise once \pref{NZ} is integrated with respect to time to give $\varrho(\eta)$ in the limit $\eta - \eta' \to \infty$. This divergence is a reflection of the breakdown of perturbative methods at late times and is related to the phenomenon of `secular' growth when $\eta - \eta'$ is finite. Part of the story told in \cite{Burgess:2022nwu} -- which also applies here -- was how these can be resummed to obtain reliable late-time behaviour, at least for super-Hubble modes. For these modes \pref{NZ} becomes approximately Markovian and can apply over a broader domain of validity than does a direct perturbative calculation of $\varrho(\eta)$.

\subsection{Gaussian property and mode mixing}

Before evaluating the environmental correlators, we pause to remark on two important simplifications -- also noted in \cite{Burgess:2022nwu} -- that occur when $\scrH_{\rm int}$ is strictly linear in the system fields, as it is in particular when restricted to interactions of the schematic form $\zeta_\sys \zeta_\env^2$ that are argued above to dominate at lowest order. In this case the right-hand side of \pref{NZ} is at most quadratic in system fields and this has two important implications: 
\begin{itemize}
\item The evolution does not mix modes with different values of $\bm{k}$, similar to free evolution.
\item An initially gaussian system state (such as the Bunch-Davies vacuum) remains gaussian. 
\end{itemize}
We elaborate about these two points below.

\subsubsection{Mode mixing}
To see the lack of mode mixing explicitly it is convenient to define the momentum-space correlation function $\mathscr{C}^{ab}_{\bm{k}}(\eta,\eta')$ using
\begin{equation}
\label{CR_FT}
C^{ab}(\eta, \eta' ; \bm{y}) =  \frac{1}{V}\sum_{\bm{k}}\; 
\mathscr{C}^{ab}_{\bm{k}}(\eta,\eta') e^{i \bm{k} \cdot \bm{y} } \,,
\end{equation}
and we treat momenta as being denumerable by placing the system in a fictitious box of volume $V$. $\mathscr{C}^{ab}_{\bm{k}}(\eta,\eta')$ is computed explicitly below and is found only to depend on the modulus $k = |\bm{k}|$ (because of the rotation invariance of the chosen environmental state). So in particular $\mathscr{C}^{ab}_{-\bm{k}}(\eta,\eta') = \mathscr{C}^{ab}_{\bm{k}}(\eta,\eta')$. 

So far as decoherence is concerned, we can omit the `tadpole' (or $\scrE$-dependent) part of the NZ equation \pref{NZ}, since it does not contribute at all to the decoherence. In terms of $\mathscr{C}^{ab}_{\bm{k}}(\eta,\eta')$ the remaining (non-tadpole) part of eq.~\pref{NZ} becomes
\begin{equation}
\label{NZmodes}
\frac{\partial \varrho}{\partial \eta}
 = - \sum_{\bm{k}} \int_{ \eta_{\mathrm{in}} }^{\eta} \exd  \eta' \;
\bigg\lbrace G(\eta) \, G(\eta') \Bigl[ \cS_{a{\bm{k}}}(\eta) ,
  \cS_{b{-\bm{k}}}(\eta')  \varrho(\eta') \Bigr]
\mathscr{C}^{ab}_{\bm{k}}(\eta,\eta') + \hbox{h.c.}
\bigg\rbrace  ,
\end{equation}
where $\cS_{a{\bm{k}}}$ is defined by (compare \pref{eq:v:Fourier}) 
\begin{equation}
\label{eq:v:FourierS}
\cS_a(\eta,\bm{x}) =\frac{1}{\sqrt{V}}\sum_{\bm{k}}\; 
\cS_{a{\bm{k}}}(\eta) \, e^{i \bm{k}\cdot\bm{x}}\,.
\end{equation}
If the reduced density matrix is prepared with different momenta uncorrelated,
\be
  \varrho = \prod_{\bm{k}} \otimes \varrho_{\bm{k}} \,,
\ee
as \pref{rhofactor} shows is true in particular for the Bunch-Davies choice, then $\varrho^{-1} \partial_\eta \varrho = \sum_{\bm{k}} \varrho_{\bm{k}}^{-1} \partial_\eta \varrho_{\bm{k}}$ with
\begin{equation}
\label{NZmodesrhok}
\frac{\partial \varrho_{\bm{k}}}{\partial \eta}
 = -   \int_{ \eta_{\mathrm{in}} }^{\eta} \exd  \eta' \;
\bigg\lbrace G(\eta) \, G(\eta') \Bigl[ \cS_{a{\bm{k}}}(\eta) ,
  \cS_{b{-\bm{k}}}(\eta')  \varrho_\bmk(\eta') \Bigr]
\mathscr{C}^{ab}_{\bm{k}}(\eta,\eta') + \hbox{h.c.}
\bigg\rbrace  \,,
\end{equation}
evolving independently for each $\bm{k}$, as claimed.

\subsubsection{System gaussianity}
\label{sssec:Gaussianity}

The second consequence of the Nakajima-Zwanzig equation being at most quadratic in system fields is the persistence this implies for initially gaussian system statistics. This in turn allows us to draw two very useful conclusions.

First, gaussianity allows a very explicit demonstration that the purity evolution one finds is ultimately the same when computed using the $\zeta$ variables or the $v$ variables (despite the coefficients appearing in the Nakajima-Zwanzig equation differing for these two representations). 

To see how, it is convenient to start by adopting the canonical field variable $v$ (rather than $\zeta$, say) and to  follow \cite{Martin:2018zbe} by breaking its Fourier components $v_{-\bm{k}}$ into real and imaginary parts 
\begin{equation} \label{alphaRI}
v_{\bm{k}}(\eta) =: \tfrac{1}{\sqrt2} \Bigl[ v^\ssR_{\bm{k}}(\eta)
+i \, v^\ssI_{\bm{k}}(\eta) \Bigr] \,,
\end{equation}
for which $v_{-\bm{k}} = v^\dagger_{\bm{k}}$ implies $v^{\ssR}_{\bm{k}} = v^{\ssR}_{-\bm{k}}$ while $v^{\ssI}_{\bm{k}} =
-v^{\ssI}_{-\bm{k}}$. These real components evolve separately under linear evolution and this evolution is identical provided the Hamiltonian is invariant under reflections in $\bm{k}$, which is true in particular if the physics involved is parity invariant or if it is invariant under arbitrary rotations (as is true here). This makes it convenient to treat the system as if it were a single real field, $\w_{\bm{k}} = \w^\dagger_{\bm{k}}$ for {\it all} $\bm{k}$ and then identify $\sqrt2 \; v^\ssR_{\bm{k}} = \w_{\bm{k}}+\w_{-\bm{k}}$ and $\sqrt2 \; v^\ssI_{\bm{k}} = \w_{\bm{k}}-\w_{-\bm{k}}$ respectively as its even and odd parts under reflections of $\bm{k}$, since this simplifies the notation by allowing us to use real variables but drop the superscripts `R' and `I' on the fields. A similar story also
applies for the canonical momentum field, whose real Fourier components we similarly denote $\p_{\bm{k}}$.

In terms of these variables the Gaussian nature of the reduced density matrix boils down to the statement that its matrix elements can be written
\be 
\label{eq:rho:Gaussian}
\left\langle \w_{\bm{k},1} \right\vert \varrho_{\bm{k}}
\left\vert \w_{\bm{k},2} \right\rangle =
\sqrt{\frac{ \mathrm{Re} \, \mfa_k -  \mfc_k }{\pi}} \; \exp\left(-\frac{\mfa_k}{2}
  \, \w^2_{\bm{k},1}  -\frac{\mfa^*_k}{2}
 \, \w^2_{\bm{k},2} 
+ \mfc_k \, \w_{\bm{k},1} \w_{\bm{k},2} \right)\,,
\ee
for some choice of time-dependent functions $\mfa_k(\eta)$ and $\mfc_k(\eta)$. As written, this state is normalised to satisfy $\mathrm{Tr}(\varrho_{\bm{k}}) = 1$ and the requirement $\varrho_{\bm{k}}^\dagger = \varrho_{\bm{k}}$ further implies $\mfc_k$ is real. From this point of view the purpose of the Nakajima-Zwanzig equation is to determine the time-dependence of the functions $\mfa_k(\eta)$ and $\mfc_k(\eta)$. 

Because the state is Gaussian the coefficients $\mfa_k$ and $\mfc_k$ are completely determined by the equal-time two-point functions, through the formulae
\bea \label{Pcorrdefs}
&&\left\langle  \w_{\bm{k}} \w_{\bm{k}'}  \right\rangle
= P_{vv}(k) \, \delta_{\bm{k},\bm{k}'} \,, \quad
 \left\langle  \p_{\bm{k}}  \p_{\bm{k}'}   \right\rangle
 = P_{pp}(k) \, \delta_{\bm{k}.\bm{k}'}, \nn\\ 
 &&\qquad\qquad \hbox{and} \quad
 \left\langle \w_{\bm{k}} {\p}_{\bm{k}'}  \right\rangle
 =  \left[P_{v p}(k)+\frac{i}{2}\right] \, 
\delta_{\bm{k},\bm{k}'}  \,,
\eea
where the quantities $P_{vv}(k)$, $P_{vp}(k)$ and $P_{pp}(k)$ are given by
\be 
P_{vv}(k)=\frac{1}{2 \left[\mathrm{Re} \, \mfa_k -\mfc_k\right]}\, ,\quad
P_{vp}(k)=-\frac{\mathrm{Im} \,\mfa_k}
{2 \left[\mathrm{Re} \, \mfa_k -\mfc_k\right]} \quad\hbox{and} \quad
P_{pp}(k)=\frac{\left\vert \mfa_k\right\vert^2-\mfc_k^2}
{2 \left[\mathrm{Re} \, \mfa_k -\mfc_k\right]}\, .
\ee
In terms of these the original complex correlators are given by
\be
  \left \langle v_\bmk^\dagger v_\bmk \right\rangle = P_{vv}(k) + P_{vv}(-k) \,, \qquad
  \left \langle p_\bmk^\dagger p_\bmk \right\rangle = P_{pp}(k) + P_{pp}(-k) \,,
\ee
and so on.

In particular, the state's purity -- {\it c.f.}~eq.~\pref{puritydef} -- is given in terms of these by \cite{Serafini:2003ke, Grain:2019vnq, Colas:2021llj,Martin:2021znx}
\be 
\label{purityk}
\gamma_{\bm{k}}(\eta) :=
\mathrm{Tr}\left[ {\varrho}^2_{\bm{k}} (\eta)\right]   = \sqrt{\frac{\mathrm{Re} \, \mfa_k -\mfc_k}{\mathrm{Re} \, \mfa_k +\mfc_k} } \,,
\ee
and this allow us to track the effects of field redefinitions on the purity, and in particular how the result changes if computed using the field $v(\bm{x},\eta)$ rather than $\zeta(\bm{x},\eta)$. Using the definition $v = \mfz(\eta) \, \zeta$ -- {\it c.f.}~eq.~\pref{eq:defzeta} -- shows that the coefficients $\hat\mfa_k$ and $\hat\mfc_k$ appearing in the density matrix
\be 
\label{eq:rho:Gaussianzeta}
\left\langle \tilde\zeta_{\bm{k},1} \right\vert \varrho_{\bm{k}}
\left\vert \tilde\zeta_{\bm{k},2} \right\rangle =
\sqrt{\frac{ \mathrm{Re} \,\hat\mfa_k -  \hat \mfc_k }{\pi}} \; \exp\left(-\frac{\hat \mfa_k}{2}
  \, \tilde\zeta^2_{\bm{k},1}  -\frac{\hat \mfa^*_k}{2}
 \, \tilde\zeta^2_{\bm{k},2} 
+ \hat \mfc_k \, \tilde\zeta_{\bm{k},1} \tilde\zeta_{\bm{k},2} \right)\,,
\ee
are related to $\mfa_k$ and $\mfc_k$ defined in \pref{eq:rho:Gaussian} by $\hat\mfa_k(\eta) = \mfz(\eta) \, \mfa_k(\eta)$ and $\hat\mfc_k(\eta) = \mfz(\eta) \, \mfc_k(\eta)$. Eq.~\pref{purityk} shows that this rescaling of $\mfa_k$ and $\mfc_k$ completely drops out of the purity.

A similar conclusion also follows for the integration by parts (or canonical transformation) that takes us from the squeezed Hamiltonian \pref{squeezedfree} to the Harmonic oscillator form \pref{freeHclassicalx}, since the change of state that this implies amounts to a shift only in the imaginary part of $\mfa_k$. Eq.~\pref{purityk} also shows that such a shift does not affect the purity. 

We see from these arguments that we are free to compute the purity using the variables that are most convenient, secure in the knowledge that the result is the same provided we include all terms that are relevant at leading semiclassical order (which was needed to conclude the evolution predicted by the Nakajima-Zwanzig equation remains gaussian). These arguments do {\it not} guarantee that the calculations are equally simple of course, and we find below that the emergence of the Markovian nature of the evolution is easier to see using the variable $\zeta$ rather than $v$.

The second useful implication of gaussian purity comes when we wish to resum our perturbative calculations at very late time, which we do not do here but discuss briefly in \S\ref{sec:Conclusions}. 

\section{Super-Hubble evolution and the Markovian limit}
\label{sec:markstrat}

We next compute the implications of the Nakajima-Zwanzig equation \pref{NZmodes} for the evolution of $\varrho_{\bm{k}}(\eta)$, focussing on the super-Hubble regime $|k \eta| \ll 1$.
Part of this story involves the extent to which this evolution becomes approximately Markovian, so in this section we start with a discussion of the $\eta$- and $\eta'$-dependence of the right-hand side of \pref{NZmodes} in order to identify the domain of validity of the Markovian approximation. 

\subsection{TCL2 form}

Our strategy moving forward starts by making the dependence on $\eta'$ of the right-hand side of the Nakajima-Zwanzig equation as explicit as possible so that the size of any non-Markovian evolution can be explicitly evaluated, since only then can we assess the price of the Markovian limit within a \emph{controlled} approximation scheme. 

\subsubsection{Field evolution}
\label{subsec:convfieldvars}

Identifying the $\eta'$ dependence hidden in the field variables evaluated at $\eta^{\prime}$ is a simple matter since we work in the interaction picture where the fields and their conjugate momenta evolve according to the free Hamiltonian given in eq.\eqref{freeHclassical}. It is therefore captured by expanding the fields at $\eta'$ in terms of the creation and annihilation operators $c_{\bm{k}}$ using \pref{vhatk} and then re-expressing the $c_{\bm{k}}$'s in terms of the fields evaluated at $\eta$ (again using \pref{vhatk}). 

The resulting evolution is then given explicitly in terms of mode functions. For the canonical fields this implies:   
\begin{eqnarray}
\label{eq:etaprimeToetav}
 v_{-\bmk }(\eta^{\prime})  &=&   \scrW_k (\eta^{\prime}, \eta) \, v_{-\bmk }(\eta)+ \scrX_k (\eta^{\prime}, \eta) \,  p_{-\bmk }(\eta)\,,\nn\\
 p_{-\bmk }(\eta^{\prime})  &=&  \scrZ_k (\eta^{\prime}, \eta) \, v_{-\bmk }(\eta)+ \scrY_k (\eta^{\prime}, \eta) \,  p_{-\bmk }(\eta) \,,
\end{eqnarray}
with the kernels $\scrW_k (\eta^{\prime}, \eta), \ldots , \scrZ_k (\eta^{\prime}, \eta)$ defined by
\begin{eqnarray}
\label{eq:etaprimeToeta2v}
 \scrW_k (\eta^{\prime}, \eta) &:=&  i\   \Bigl[ u_k^*(\eta^{\prime}) Du_k(\eta)-u_k(\eta^{\prime}) Du_k^{*}(\eta)\Bigr] \nn \\
 \scrX_k (\eta^{\prime}, \eta) &:=&  i\ \Bigl[ u_k(\eta^{\prime}) u_k^{*}(\eta)-u_k^*(\eta^{\prime}) u_k(\eta)\Bigr]\\
 \scrY_k (\eta^{\prime}, \eta) &:=& i\   \Bigl[ Du_k(\eta^{\prime}) u_k^{*}(\eta)-Du_k^{*}(\eta^{\prime}) u_k(\eta)\Bigr]\nn\\
 \scrZ_k (\eta^{\prime}, \eta) &:=&i\   \Bigl[ Du_k^{*}(\eta^{\prime}) Du_k(\eta)-Du_k(\eta^{\prime}) Du_k^{*}(\eta)\Bigr]\,,\nn
\end{eqnarray}
and $D$ as defined in \pref{Ddef}. 
For example in the case of massless fields in de Sitter $u_k(\eta)$ is given explicitly by eq.~\pref{eq:deSitter:BD:vk}. Notice that $\scrX_k \to 0$ as $\eta' \to \eta$ (as does $\scrZ_k$) whereas the canonical normalization of the modes ensures $\scrW_k \to +1$ and $\scrY_k \to - 1$ in this limit. The corresponding expressions for $\zeta_\bmk(\eta) = v_\bmk(\eta) /\mfz(\eta)$ and $\mfp_\bmk(\eta) = p_\bmk(\eta) \, \mfz(\eta)$ therefore become
\begin{eqnarray}
\label{eq:etaprimeToeta}
\zeta_{-\bmk }(\eta^{\prime})  &=&  \cW_k (\eta^{\prime}, \eta) \, \zeta_{-\bmk }(\eta)+\cX_k (\eta^{\prime}, \eta) \, \mfp_{-\bmk }(\eta)\,,\nn\\
\mfp_{-\bmk }(\eta^{\prime})  &=&  \cZ_k (\eta^{\prime}, \eta) \, \zeta_{-\bmk }(\eta)+\cY_k (\eta^{\prime}, \eta) \, \mfp_{-\bmk }(\eta) \,,
\end{eqnarray}
with the kernels $\cW_k (\eta^{\prime}, \eta), \ldots \cZ_k (\eta^{\prime}, \eta)$ now given by
\begin{eqnarray}
\label{eq:etaprimeToeta2}
 \cW_k (\eta^{\prime}, \eta) &:=&  i\ \frac{\mfz(\eta)}{\mfz(\eta')} \Bigl[ u_k^*(\eta^{\prime}) Du_k (\eta)-u_k(\eta^{\prime}) D u_k^{*}(\eta)\Bigr] 
 =   i\ \mfz^2(\eta) \Bigl[ \hat u_k^*(\eta^{\prime}) \hat u_k^{\prime}(\eta)- \hat u_k(\eta^{\prime}) \hat u_k^{* \prime}(\eta)\Bigr] \nn \\
 \cX_k (\eta^{\prime}, \eta) &:=&  \frac{i}{\mfz(\eta') \, \mfz(\eta)} \Bigl[ u_k(\eta^{\prime}) u_k^{*}(\eta)-u_k^*(\eta^{\prime}) u_k(\eta)\Bigr]  =   i  \Bigl[ \hat u_k(\eta^{\prime}) \hat u_k^{*}(\eta)- \hat u_k^*(\eta^{\prime}) \hat u_k(\eta)\Bigr]\\
 \cY_k (\eta^{\prime}, \eta) &:=&  i\ \frac{\mfz(\eta')}{\mfz(\eta)}  \Bigl[ Du _k(\eta^{\prime}) u_k^{*}(\eta)-Du_k^{*}(\eta^{\prime}) u_k(\eta)\Bigr] 
 =   i\ \mfz^2(\eta')  \Bigl[ \hat u' _k(\eta^{\prime}) \hat u_k^{*}(\eta)-\hat u_k^{*\prime}(\eta^{\prime}) \hat u_k(\eta)\Bigr]\nn\\
 \cZ_k (\eta^{\prime}, \eta) &:=&i\  \mfz(\eta) \,\mfz(\eta^{\prime}) \ \Bigl[ Du_k^{*}(\eta^{\prime}) Du_k(\eta)- Du_k(\eta^{\prime}) Du_k^{*}(\eta)\Bigr] \nn\\
 &&\qquad\qquad\qquad\qquad\qquad\qquad= i\  \mfz^2(\eta) \,\mfz^2(\eta^{\prime}) \ \Bigl[ \hat u_k^{*\prime}(\eta^{\prime}) \hat u_k^{\prime}(\eta)- \hat u_k^{\prime}(\eta^{\prime}) \hat u_k^{*\prime}(\eta)\Bigr]\,.\nn
\end{eqnarray}
The two equalities give the expressions for both the $\zeta$ and $v$ mode functions \pref{modev} and \pref{eq:deSitter:BD:vk}. Again $\cX_k, \cZ_k \to 0$ as $\eta' \to \eta$ and canonical normalization ensures $\cW_k \to +1$ and $\cY_k \to - 1$ in this limit. 

Expression \pref{eq:etaprimeToeta} achieves the goal of explicitly exhibiting the $\eta'$ dependence previously hidden in the fields themselves, with the price paid being the various kernels in eq.\pref{eq:etaprimeToeta2}.

\subsubsection{Density matrix evolution}
\label{subsec:convreddens}

We next make explicit the $\eta^{\prime}$ dependence in $\varrho(\eta^{\prime})$. This can be done by simply Taylor expanding $\varrho(\eta^{\prime})$ around $\eta^{\prime}=\eta$, and (to start with) keeping {\it all} the terms:
\begin{equation} \label{TaylorRho}
\varrho(\eta^{\prime})=\sum_{n=0}^{\infty} \frac{\partial_{\eta}^n \varrho(\eta)}{n!} (\eta^{\prime}-\eta)^n \,.
\end{equation}
The price paid here is the acquired dependence on all of the derivatives of $\varrho$ evaluated at $\eta$. 

As emphasized in the introduction, we do {\it not} simply truncate this series and hope for the best. We also do not make use of the argument that these higher derivatives are themselves higher order in the perturbative coupling (and so can automatically be neglected at the order we work in the semiclassical perturbative expansion). As stated earlier, this argument becomes suspicious precisely when $\eta - \eta'$ is large (where perturbation theory inevitably breaks down), so its use would obstruct our quest for a controlled approximation scheme valid at late times. We seek a different reason for neglecting these terms.

\subsubsection{TCL2 evolution}

We next assemble these expressions to rewrite the Nakajima-Zwanzig equation \pref{NZmodes} in an equivalent so-called `time convolution-less' (TCL) form \cite{BreuerPetruccione}. To do so we insert the field evolution of eq.~\pref{eq:etaprimeToeta} and the Taylor expansion of eq.~\pref{TaylorRho} into eq.~\pref{NZmodes}, using the explicit time-conversion kernels given in eq.\pref{eq:etaprimeToeta2}. This leads to the `TCL2' result:\footnote{TCL2 denotes the form of the TCL equation at second-order in perturbation theory.} 
\bea
\label{eq:TCL2}
\partial_{\eta} \varrho_{\bmk } &=& -\sum_{n=0}^{\infty} \frac{1}{n!} \Biggl \{  \mathcal{I}^{a \zeta}_{kn}(\eta;\eta_{\rm in})\Bigl[\cS_{a\bmk}(\eta),\ \zeta_{-\bmk}(\eta)\ \partial^n_{\eta}\varrho(\eta)\Bigr] \nn\\
&&\qquad\qquad\qquad\qquad   \vphantom{\frac12}  +\mathcal{I}^{a \mfp}_{k n}(\eta;\eta_{\in})\Bigl[\cS_{a\bmk}(\eta),\ \mfp_{-\bmk}(\eta)\ \partial^n_{\eta}\varrho(\eta)\Bigr] +{\rm h.c.}\Biggr\},
\eea
where the $\eta'$ integrals now have an explicit integrand:
\begin{subequations}
\label{eq:pvkernels}
\begin{align}
\mathcal{I}^{a \zeta}_{k n}(\eta;\eta_{\in}) &= {G}(\eta)  \int_{\eta_{\in}}^{\eta} \exd\eta^{\prime}\  {G}(\eta^{\prime}) \, \mathcal{K}^\zeta_{b k}(\eta^{\prime},\eta)\mathscr{C}^{ab}_{\bm{k}}(\eta,\eta') \, (\eta^{\prime}-\eta)^n\\
\mathcal{I}^{a \mfp}_{k n}(\eta;\eta_{\in}) &=  {G}(\eta) \int_{\eta_{\in}}^{\eta} \exd\eta^{\prime}\  {G}(\eta^{\prime})  \,\mathcal{K}^\mfp_{b k}(\eta^{\prime},\eta)\mathscr{C}^{ab}_{\bm{k}}(\eta,\eta') \, (\eta^{\prime}-\eta)^n. 
\end{align}
\end{subequations}

The kernels $\mathcal{K}^{\zeta,\mfp}_{b k}(\eta^{\prime},\eta)$ in \pref{eq:pvkernels} are obtained by decomposing the operator $\cS_{b\, - \bm{k}}(\eta')$ in terms of $\zeta_{-\bm{k}}(\eta')$ and $\mfp_{-\bm{k}}(\eta')$ and choosing the appropriate choice of the kernel in eq.~\pref{eq:etaprimeToeta2} that applies. $\mathcal{K}^\zeta_{b k}$ corresponds to the kernel appropriate for the term involving $\zeta_{-\bm{k}}(\eta)$ on the right-hand side of \pref{eq:etaprimeToeta} -- and so involves $\cW_k$ for terms in $\cS_{a\,{-\bm{k}}}(\eta')$ involving $\zeta_{-\bm{k}}(\eta')$ or $\cZ_k$ for terms in $\cS_{a\,{-\bm{k}}}(\eta')$ involving $\mfp_{-\bm{k}}(\eta')$.  $\mathcal{K}^\mfp_{b k}$ similarly corresponds to contributions involving $\mfp_{-\bm{k}}(\eta)$ on the right-hand side of \pref{eq:etaprimeToeta} -- and so involves $\cX_k$ or $\cY_k$ depending on whether one focusses on $\zeta_{-\bm{k}}(\eta')$ or $\mfp_{-\bm{k}}(\eta')$ in $\cS_{a\,-\bm{k}}(\eta')$. 

Further grouping the results to isolate the contributions of $\zeta_{\bm{k}}$ and $\mfp_{\bm{k}}$ in $\cS_{a\,\bm{k}}$ leads to the final form we use when assessing the size of interactions when evolving:
\bea
\label{eq:TCL2a}
\partial_{\eta} \varrho_{\bmk }(\eta) &=& -\sum_{n=0}^{\infty} \frac{1}{n!} \Biggl \{  \mathcal{J}^{\zeta \zeta}_{kn}(\eta;\eta_{in})\Bigl[\zeta_{\bmk}(\eta),\ \zeta_{-\bmk}(\eta)\ \partial^n_{\eta}\varrho(\eta)\Bigr] + \mathcal{J}^{\zeta \mfp}_{kn}(\eta;\eta_{in})\Bigl[\zeta_{\bmk}(\eta),\ \mfp_{-\bmk}(\eta)\ \partial^n_{\eta}\varrho(\eta)\Bigr] \nn\\
&&\; \vphantom{\frac12} +\mathcal{J}^{\mfp \zeta}_{k n}(\eta;\eta_{\in})\Bigl[\mfp_{\bmk}(\eta),\ \zeta_{-\bmk}(\eta)\ \partial^n_{\eta}\varrho(\eta)\Bigr] +\mathcal{J}^{\mfp \mfp}_{k n}(\eta;\eta_{\in})\Bigl[\mfp_{\bmk}(\eta),\ \mfp_{-\bmk}(\eta)\ \partial^n_{\eta}\varrho(\eta)\Bigr] +{\rm h.c.}\Biggr\},\nn\\
\eea
where the $\eta'$ integrals now become:
\begin{subequations}
\label{eq:pvkernels}
\begin{align}
\mathcal{J}^{\zeta \zeta}_{k n}(\eta;\eta_{\in}) &= {G}(\eta)  \int_{\eta_{\in}}^{\eta} \exd\eta^{\prime}\  {G}(\eta^{\prime}) \, \mathcal{T}^{\zeta\zeta}_{k}(\eta^{\prime},\eta)  \, (\eta^{\prime}-\eta)^n \\
\mathcal{J}^{\zeta \mfp}_{k n}(\eta;\eta_{\in}) &=  {G}(\eta) \int_{\eta_{\in}}^{\eta} \exd\eta^{\prime}\  {G}(\eta^{\prime})  \,\ \mathcal{T}^{\zeta\mfp}_{k}(\eta^{\prime},\eta) \, (\eta^{\prime}-\eta)^n \\
\mathcal{J}^{\mfp \zeta}_{k n}(\eta;\eta_{\in}) &=  {G}(\eta) \int_{\eta_{\in}}^{\eta} \exd\eta^{\prime}\  {G}(\eta^{\prime})  \, \mathcal{T}^{\mfp\zeta}_{k}(\eta^{\prime},\eta) \, (\eta^{\prime}-\eta)^n \\
\mathcal{J}^{\mfp \mfp}_{k n}(\eta;\eta_{\in}) &=  {G}(\eta) \int_{\eta_{\in}}^{\eta} \exd\eta^{\prime}\  {G}(\eta^{\prime})  \, \mathcal{T}^{\mfp\mfp}_{k}(\eta^{\prime},\eta) \, (\eta^{\prime}-\eta)^n. 
\end{align}
\end{subequations}
The grouped correlators appearing here are given by 
\begin{equation}
\label{eq:zetazeta}
\begin{split}
\mathcal{T}_k^{\zeta\zeta}(\eta,\eta^{\prime}) &=\cW_k(\eta,\eta^{\prime})\left\langle \left[\cE^{\zeta}+\cE^{\zeta i} \left(-i\hat{\bmk}_i k\right)\right]_{\eta} \left[\cE^{\zeta}+\cE^{\zeta i} \left(i\hat{\bmk}_i k\right)\right]_{\eta^{\prime}} \right\rangle\\
&\qquad\qquad + \cZ_k(\eta,\eta^{\prime}) \left\langle \left[\cE^{\zeta}+\cE^{\zeta i} \left(-i\hat{\bmk}_i k\right)\right]_{\eta} \left[ \cE^{\mfp}+\cE^{\mfp i} \left(-i\frac{\hat{\bmk}_i }{k}\right)\right]_{\eta^{\prime}} \right\rangle ,
\end{split}
\end{equation}
\begin{equation}
\label{eq:zetap}
\begin{split}
\mathcal{T}_k^{\zeta\mfp}(\eta,\eta^{\prime}) &=\cX_k(\eta,\eta^{\prime})\left\langle \left[\cE^{\zeta}+\cE^{\zeta i} \left(-i\hat{\bmk}_i k\right)\right]_{\eta} \left[\cE^{\zeta}+\cE^{\zeta i} \left(i\hat{\bmk}_i k\right)\right]_{\eta^{\prime}} \right\rangle\\
&\qquad\qquad + \cY_k(\eta,\eta^{\prime}) \left\langle \left[ \cE^{\zeta}+\cE^{\zeta i} \left(-i\hat{\bmk}_i k\right)\right]_{\eta} \left[\cE^{\mfp}+\cE^{\mfp i} \left(-i\frac{\hat{\bmk}_i }{k}\right)\right]_{\eta^{\prime}} \right\rangle ,
\end{split}
\end{equation}
\begin{equation}
\label{eq:pzeta}
\begin{split}
\mathcal{T}_k^{\mfp\zeta}(\eta,\eta^{\prime}) &=\cW_k(\eta,\eta^{\prime})\left\langle \left[\cE^{\mfp}+\cE^{\mfp i} \left(i\frac{\hat{\bmk}_i }{k}\right)\right]_{\eta}  \left[\cE^{\zeta}+\cE^{\zeta i} \left(i\hat{\bmk}_i k\right)\right]_{\eta^{\prime}} \right\rangle\\
&\qquad\qquad + \cZ_k(\eta,\eta^{\prime}) \left\langle\left[\cE^{\mfp}+\cE^{\mfp i} \left(i\frac{\hat{\bmk}_i }{k}\right)\right]_{\eta}\left[\cE^{\mfp}+\cE^{\mfp i} \left(-i\frac{\hat{\bmk}_i }{k}\right)\right]_{\eta^{\prime}} \right\rangle ,
\end{split}
\end{equation}
and
\begin{equation}
\label{eq:pp}
\begin{split}
\mathcal{T}_{\mfp\mfp}(\eta,\eta^{\prime}) &= \cY_k(\eta,\eta^{\prime}) \left\langle \left[\cE^{\mfp}+\cE^{\mfp i} \left(i\frac{\hat{\bmk}_i }{k}\right)\right]_{\eta} \left[\cE^{\mfp}+\cE^{\mfp i} \left(-i\frac{\hat{\bmk}_i }{k}\right)\right]_{\eta^{\prime}} \right\rangle\\
&\qquad\qquad + \cX_k(\eta,\eta^{\prime}) \left\langle \left[\cE^{\mfp}+\cE^{\mfp i} \left(i\frac{\hat{\bmk}_i }{k}\right)\right]_{\eta} \left[\cE^{\zeta}+\cE^{\zeta i} \left(i\hat{\bmk}_i k\right)\right]_{\eta^{\prime}}  \right\rangle.
\end{split}
\end{equation}
The various factors involving $\hat{\bmk}$ come about from converting the spatial derivatives acting on the system operators; the difference in sign between operators evaluated at $\eta$ versus $\eta^{\prime}$ come about because the latter involve the complex conjugates of the modes. 

Now comes the main point: in the sections to follow we claim that in the superhorizon limit where $z=-k\eta\ll1$ we can neglect the effects of the kernels $\mathcal{J}^{rs}_{k n}(\eta;\eta_{in})$ for all $r$ and $s$ and for all $n\geq 1$ because these are all suppressed by at least one power of $z$ compared to the leading result $\mathcal{J}^{rs}_{k \, 0}(\eta;\eta_{in})$. We also find that $\mathcal{J}^{\zeta\zeta}_{k \, 0}(\eta;\eta_{in})$ is less suppressed than the other three $\mathcal{J}^{rs}_{k \, 0}(\eta;\eta_{in})$ by at least one power of $z$, and this makes it dominate the eigenvalues in the Markovian evolution that eventually emerges. The leading forms for all four of the $\mathcal{J}^{rs}_{k \, 0}(\eta;\eta_{in})$ turn out to compete in the predictions for the purity evolution. 

The same counting of powers of $z$ does not hold when computing using $v$, because $v$ does not freeze on super-Hubble scales in the way $\zeta$ does. This does not mean that the purity evolution differs for calculations with $v$ and $\zeta$, but it does make the calculations using $v$ more difficult to do due to the need to include more terms to obtain the same answer (see Appendix \ref{App:Comparison} for an explicit example).

\subsection{Late times and the Markovian limit}

The first step in identifying the dominant terms in \pref{eq:TCL2a} is a precise statement of the integration contour. We follow \cite{Maldacena:2002vr} and take this integration contour to be deformed into the complex plane using $\eta' = \eta' (1 - i \varepsilon) =  \eta^{\prime}+i\varepsilon \left|\eta^{\prime}\right|$ where $\varepsilon > 0$ is taken to zero at the end of the day. This serves two separate purposes. It firstly provides the customary small imaginary part that is required to make the Wightman function well-defined for finite times, by suppressing the contribution of very high-energy intermediate states. But having the contour move further and further into the complex plane in the remote past also projects onto the adiabatic vacuum state in this limit, ensuring the initial state at $\eta_{\in} \to - \infty$ is prepared in the Bunch-Davies vacuum.  

The next conceptual issue is the singular nature of the correlation functions $\mathscr{C}^{ab}_{\bm{k}}(\eta,\eta')$ as $\eta^{\prime}\rightarrow \eta$ since this causes the kernels $\mathcal{I}^{a \mfp}_{k n}(\eta;\eta_{in})$ and $\mathcal{I}^{a \zeta}_{k n}(\eta;\eta_{in})$ to diverge. Equivalently, the corresponding singular behaviour in $\cT^{\zeta\zeta}_{k}(\eta,\eta')$, $\cT^{\mfp \zeta}_k(\eta,\eta')$, $\cT^{\zeta \mfp}_k(\eta,\eta')$ and $\cT^{\mfp\mfp}_k(\eta, \eta')$ cause the integrals $\cJ^{\zeta\zeta}_{kn}(\eta,\eta_{\in})$, $\cJ^{\zeta \mfp}_{kn}(\eta,\eta_{\in})$, $\cJ^{\mfp \zeta}_{kn}(\eta,\eta_{\in})$ and $\cJ^{\mfp\mfp}_{kn}(\eta,\eta_{\in})$ to diverge. These divergences must be regulated in order to understand their behaviour in the small $k\eta$ limit, and we do so using the $\eta^{\prime}$ integration contour itself, since having $\varepsilon > 0$ regulates these diverges. If not renormalized they show up again as singularities when the limit $\varepsilon \to 0$ is taken at the end of the calculation. These calculations might profit by using more sophisticated analytic regularization schemes \cite{Premkumar:2021mlz, Beneke:2023wmt}.

\subsubsection{Scaling arguments and power counting}

It is useful to rewrite eqs.~\pref{eq:TCL2a} and \pref{eq:pvkernels} in terms of dimensionless variables so that the dependence on the scales of the problem is made explicit. We do so by scaling out $k$ as required on dimensional grounds, trading $\eta$ and $\eta'$ for 
\be
  z := -k\eta \qquad \hbox{and} \qquad z^{\prime} := -k\eta^{\prime} \qquad \hbox{and so} \qquad w := k(\eta-\eta^{\prime})=z^{\prime}-z \,.
\ee 

With these choices the dimensionless mode functions obtained from eq.~\pref{eq:deSitter:BD:vk} become  
\be
 \tilde{u}_k(z) := (1-i z) \, e^{i z} \qquad \hbox{so} \qquad \hat u_k(\eta) = \frac{i H}{\sqrt{2 k^3 \slrl}} \; \tilde u_{{k}}(z) \,,
\ee 
and the natural dimensionless form for the kernels in eq.~\pref{eq:etaprimeToeta2} (evaluated in de Sitter space) is 
\begin{eqnarray}
\label{eq:etaprimeToeta2dimless}
 \widetilde{\cW}_k(z^{\prime},z) &:=& \frac{i}{2}\Bigl[ \tilde{u}_k^*(z^{\prime}) \partial_z\tilde{u}_k(z)-\tilde{u}_k(z^{\prime}) \partial_z\tilde{u}^*_k(z)\Bigr]   = z^2 \cW_k (\eta^{\prime}, \eta)  \,, \nn \\
 \widetilde{\cX}_k (z^{\prime}, z) &:=& \frac{i}{2}\Bigl[ \tilde{u}_k^*(z^{\prime}) \tilde{u}_k(z)-\tilde{u}_k(z^{\prime}) \tilde{u}^*_k(z)\Bigr] = 2\slrl k^3 \left(\frac{\Mp}{H}\right)^2 \cX_k (\eta^{\prime}, \eta) \,, \\
\widetilde{\cY}_k(z^{\prime},z) &:=& \frac{i}{2}\Bigl[\partial_{z^{\prime}}\tilde{u}_k(z^{\prime}) \tilde{u}^*_k(z)-\partial_{z^{\prime}}\tilde{u}_k^*(z^{\prime}) \tilde{u}_k(z)\Bigr] =   (z')^2 \cY_k(\eta^{\prime},\eta) \,,  \nn\\
\widetilde{\cZ}_k(z^{\prime},z)  &:=& \frac{i}{2}\Bigl[\partial_{z^{\prime}} \tilde{u}_k^*(z^{\prime}) \partial_z \tilde{u}_k(z)-\partial_{z^{\prime}} \tilde{u}_k(z^{\prime}) \partial_z \tilde{u}_k^*(z) \Bigr] = \frac{z^2 (z')^2}{2\slrl k^3} \left(\frac{H}{\Mp}\right)^2 \cZ_k (\eta^{\prime}, \eta)  \,.\nn
\end{eqnarray}

Collecting the powers of $k$, $\slrl$ and $H\slash \Mp$ coming from the integrand and couplings $G(\eta)$ and $G(\eta^{\prime})$ in eqs.~\pref{eq:pvkernels} then allows the integrals $\cJ^{rs}_{kn}$ (with $r,s=1,2$ representing $\zeta$ and $\mfp$) to be written 
\bea \label{Jdimless}
\mathcal{J}^{rs}_{k n}(z;z_{\in}) 
 &=& k^{3-n}\int_{-i\varepsilon z}^{z_{\in}-z-i\varepsilon z_{\in}}\ \exd w\   \left(\frac{\slrl \Mp}{H}\right)^4 \frac{(-w)^n}{z^2 (z+w)^2} \; \cT^{rs}_{k}(z^{\prime},z)  \nn\\
 &=& \tfrac14 k^{p_{rs}+3-n}\int_{-i\varepsilon z}^{z_{\in}-z-i\varepsilon z_{\in}}\  \exd w \ \slrl^2\ \frac{(-w)^n}{z^2 (z+w)^2} \; \widetilde{ \mathcal{T}}^{rs}_{k}(z^{\prime},z)  \,,
\eea 
where the dimensionless correlation function is defined by 
\begin{equation}\label{tildeTdef}
   \mathcal{T}^{rs}_{k}(\eta^{\prime},\eta) :=  \frac{k^{p_{rs}}}{(2\slrl)^2} \left(\frac{H}{\Mp}\right)^4   \widetilde{\cT}^{rs}_{k}(z^{\prime},z) \,,
\end{equation}
with the prefactor coming from the overall scaling dimension $k^{p_{rs}}$ of the correlator in momentum space together with the factors associated with the four mode functions involved in constructing the correlator. For example $p_{11} = 1$ for the correlator of $\cE^\zeta = (\partial \zeta)^2 + \cdots$ since its momentum-space components have dimension (mass)${}^{1/2}$.

\subsubsection{Markovian limit}

We now argue that all of the terms in $\cJ^{rs}_{kn}$ involving $n \geq 1$ are suppressed relative to the $n=0$ term in the super-Hubble regime by at least one power of $z \ll 1$. This is ultimately why the evolution for super-Hubble modes is Markovian, and it has its roots in the fact that the super-horizon modes of the variable $\zeta$ are on very general grounds frozen, making them evolve very slowly compared with the underlying correlation scale of the environment (which is set by $H$). 

In order to make this argument we need to know something about the form of the correlation functions $\cT^{rs}_{kn}(z',z)$, which in turn requires knowing the behaviour of the environmental correlators $\scrC^{ab}_\bmk(z,z')$ appearing in \pref{eq:pvkernels}. As is shown in the next section, when evaluated in the Bunch-Davies vacuum these have the generic form
\begin{equation}
\label{eq:genericorrform}
\scrC_k^{a b}(z,z') = \frac{e^{-2 i \kappa w}}{8\pi^2} \Bigl[ \cC_0^{a b}(w,z)+e^{-i w} \cC_1^{a b}(w,z)\Bigr],
\end{equation}
where $\kappa=k_{\UV}\slash k$ and the coefficients $\cC_0^{a b}(w,z),\ \cC_1^{a b}(w,z)$ take the form:
\begin{equation}
\label{eq:corrcoeffs}
\cC^{a b}_i = \sum_{u=-u_{\rm max}}^{u_{\rm min}} \frac{f^{ab}_{iu}(z)}{w^u} \,.
\end{equation}

Combining this with the factor $(-w)^n$ appearing in \pref{Jdimless} -- whose roots are the factor $(\eta' - \eta)^n \partial_\eta^n \varrho$ in the expansion \pref{TaylorRho} --  the generic integrals that are required have the form 
\begin{equation}
\label{eq:genericintegrals}
I_j =   \int_{-i\varepsilon z}^{z_{\in}-z-i\varepsilon z_{\in}} \frac{\exd w}{w^j} \; e^{-i \beta w} = \frac{1}{z^{j-1}}\int^{\frac{z_{\in}}{z}(1-i\varepsilon)-1 }_{-i \varepsilon} \frac{\exd x}{x^j} \; e^{-i\beta z x} \,,
\end{equation}
where $\beta = 2\kappa$ or $\beta = 2\kappa +1$ while $j = u - n$ and the last equality scales $z$ out by changing the integration variable to $x=w\slash z$. Restricting to the case where the initial condition is set in the remote past, we use $z_{\in}\gg1$ and so the upper limit of this integral can be taken to be $\infty(1-i\varepsilon)$. In this limit small and positive $\varepsilon$ provides convergence for the integral for large $x$ and the result is dominated by its behaviour near $x =-i\varepsilon$, leading to the result
\begin{equation}
I_j\simeq \frac{i^{j-1}}{j-1} (z\varepsilon)^{1-j}+c_j,
\end{equation}
where $c_j$ is independent of $z$. Keeping in mind that $j = u-n$ (and keeping $\varepsilon\neq 0$), we see that taking $n \to n+1$ suppresses the dominant contribution by an additional power of $z$. This is true for each of the integrals, and means that we can consistently neglect the terms in the Nakajima-Zwanzig equation containing derivatives of $\varrho$ so long as additional powers of $z$ can be neglected. 

In the above discussion an important role was played by the limit $z_{\in} \to \infty$. For large but finite $z_{\in}$ we see that so long as we keep $\varepsilon>0$, terms depending on $z_{\in}$ are exponentially suppressed. This can be seen by going back to eq.~\pref{eq:genericintegrals} and noticing that the integral gives an incomplete gamma function that can be expanded for $z_{\in}\gg1$. Doing so gives
\begin{equation}
\label{eq:largezinsaling}
I_j\propto e^{-\beta z_{\in} \varepsilon}\  \frac{e^{i\beta(z-z_{\in})}}{(\beta z_{\in})^j} \,.
\end{equation}
This exponential suppression is expected; as we go back into the past, the non-zero value of $\varepsilon$ forces the environment into its true vacuum state. Similar effects were found when studying the relation between decoherence and decoupling \cite{Burgess:2024heo}, where the critical importance of the ordering of limits between $z_{\in}\rightarrow \infty$ and $\varepsilon\rightarrow 0$ was noted.

In principle this leaves the four integrals $\cJ^{\zeta\zeta}_{k0}$, $\cJ^{\zeta\mfp}_{k0}$, $\cJ^{\mfp\zeta}_{k0}$ and $\cJ^{\mfp\mfp}_{k0}$ as the dominant coefficients  respectively multiplying the terms $\left[\zeta_{\bmk}(\eta), \zeta_{-\bmk}(\eta)\varrho(\eta)\right]$, $\left[\zeta_{\bmk}(\eta), \mfp_{-\bmk}(\eta)\varrho(\eta)\right]$, $\left[\mfp_{\bmk}(\eta), \zeta_{-\bmk}(\eta) \varrho(\eta)\right]$ and $\left[\mfp_{\bmk}(\eta), \mfp_{-\bmk}(\eta)\varrho(\eta)\right]$ in the Nakajima-Zwanzig equation \pref{eq:TCL2a}. The  leading evolution for super-Hubble modes is Markovian for {\it two} reasons: 
\begin{itemize}
\item It is Markovian within the domain of validity of perturbation theory since time derivatives of $\varrho$ appearing on the right-hand side of equations like \pref{eq:TCL2a} are themselves proportional to powers of the expansion parameter $\lambda := \sqrt{\slrl} \, (H/\Mp)$.
\item It remains Markovian at late times -- even once secular effects make perturbation theory alone break down -- because small $z$ alone protects it.
\end{itemize} 

How the small parameters of the problem contribute to the evolution is most clearly displayed by rescaling to define the operators 
\be \label{OiandZdefs}
 \mathcal{O}_{1} := Z^{-1} \zeta \qquad \hbox{and} \qquad
 \mathcal{O}_{2} := Z \mfp \qquad \hbox{with} \qquad
 Z := \frac{1}{\sqrt{2\slrl k^3}} \left( \frac{H}{\Mp}\right) \,. 
\ee 
In terms of these operators, one can rescale $\mathcal{J}^{\zeta \zeta}$ to dimensionless objects using
\begin{equation} \label{curlyJdimless}
  \begin{split}
\lim_{\eta_{\in} \to - \infty} \tfrac{4}{\epsilon_1^2 k^4}  \mathcal{J}^{\zeta \zeta}_{k0}(\eta, \eta_{\mathrm{in}}) & =  \widetilde{\mathcal{J}}^{\zeta \zeta}_{k0}(z) \\
\lim_{\eta_{\in} \to - \infty}  \tfrac{8M_p^2}{\epsilon_1 k H^2} \mathcal{J}^{\zeta \mfp}_{k0}(\eta, \eta_{\mathrm{in}})  & =  \widetilde{\mathcal{J}}^{\zeta \mfp}_{k0}(z)
  \end{split}
\qquad \qquad
  \begin{split}
\lim_{\eta_{\in} \to - \infty}  \tfrac{8M_p^2}{\epsilon_1 k H^2} \mathcal{J}^{\mfp \zeta}_{k0}(\eta, \eta_{\mathrm{in}})  & =  \widetilde{\mathcal{J}}^{\mfp \zeta}_{k0}(z) \\
\lim_{\eta_{\in} \to - \infty}  \tfrac{16 k^2 M_p^4}{H^4} \mathcal{J}^{\mfp \mfp}_{k0}(\eta, \eta_{\mathrm{in}})  & =\widetilde{\mathcal{J}}^{\mfp \mfp}_{k0}(z) 
  \end{split}
\end{equation}
with which eq~\pref{eq:TCL2a} (neglecting the terms with $n\geq1$) takes the manifestly Lindblad form
\begin{equation}
\label{eq:lindbladlikeform}
\partial_{\eta} \varrho_{\bmk} = -i\Bigl[ \scrH_{\rm eff}(\eta), \varrho_{\bmk}(\eta)\Bigr] +\sum_{r,s=1}^2 h^{s r}_\bmk\left[\mathcal{O}_{\bmk,s}\varrho_{\bmk}\mathcal{O}^{\dag}_{\bmk,r}-\frac{1}{2}\left\{\mathcal{O}^{\dag}_{\bmk,r}\mathcal{O}_{\bmk,s},\varrho_{\bmk}\right\}\right] 
\end{equation}
as shown in Appendix \ref{App:LindTransport}. Here the matrix of couplings is given by:
\begin{equation}
\label{eq:decohmatrixh}
\left(\begin{array}{cc}h_\bmk^{11}  & h_\bmk^{12}  \\ h_\bmk^{21}  & h_\bmk^{22}  \end{array}\right)  = \frac{\slrl k}{8}\left(\frac{H}{\Mp}\right)^2\left(\begin{array}{cc}{2 \rm Re}[\widetilde{\cJ}^{\zeta\zeta}_{k0}] & \widetilde{\cJ}^{\zeta\mfp}_{k0} + \widetilde{\cJ}^{\mfp\zeta*}_{k0} \\ \widetilde{\cJ}^{\mfp\zeta}_{k0} +\widetilde{\cJ}^{\zeta\mfp *}_{k0} & 2 {\rm Re}[\widetilde{\cJ}^{\mfp\mfp}_{k0}] \end{array}\right) \,,
\end{equation}
and $\scrH_{\rm eff}$ contains both the environmental average of the interaction Hamiltonian \pref{Hintmom} and those parts of the second-order contributions that have Liouville form ({\it i.e.}~the commutator of something with $\varrho$), explicltly given by
\begin{eqnarray}
\scrH_{\rm eff} = \mathrm{Im}[ \mathcal{J}^{\zeta \zeta}_{k0}] \zeta^2 - \mathrm{Im}[\mathcal{J}^{\zeta \mfp}_{k0} + \mathcal{J}^{\mfp \zeta}_{k0}]  \tfrac{1}{2} \{\zeta,\mfp \} + \mathrm{Im}[\mathcal{J}^{\mfp \mfp}_{k0}] \mfp^2 \ .
\end{eqnarray} 
Because these involve Hamiltonian evolution they do not enter calculations of the rate of decoherence.

We show below by explicit calculation that $\cJ^{\zeta\zeta}_{k0}$ scales for small $z$ like $z^{-4}$ (with a divergent coefficient) while $\cJ^{\zeta\mfp}_{k0}$ and $\cJ^{\mfp\zeta}_{k0}$ scale like $z^{-3}$ and $\cJ^{\mfp\mfp}_{k0}$ scales like $z^{-2}$, showing that to leading order in $z$ only the $\cJ^{\zeta\zeta}_{k0}\, \left[\zeta_{\bmk}(\eta), \zeta_{-\bmk}(\eta)\varrho(\eta)\right]$ term dominates.

\section{Environmental correlators for $|k \eta| \ll 1$}
\label{sec:SHL}

Our remaining task is to evaluate the correlation functions $\scrC^{ab}_\bmk(\eta,\eta')$ and from these the quantities $\widetilde\cT^{rs}_{k}(z',z)$ and so compute the integrals $\widetilde\cJ^{rs}_{k0}$ appearing in \pref{eq:decohmatrixh} and thereby find the implication of \pref{eq:lindbladlikeform} for the evolution of the purity. The operational logic we follow when doing so is:
\begin{itemize}
\item Compute the various correlation functions $\scrC^{ab}_\bmk$ needed to construct $\cT^{\zeta \zeta},\ \cT^{\zeta\mfp},\ \cT^{\mfp \zeta},\\cT^{\mfp \mfp}$. When doing so we integrate using the time contour deformation $\eta^{\prime} \rightarrow \eta^{\prime}+i\varepsilon  \left|\eta^{\prime}\right|$ described above to ensure we project reliably onto the Bunch-Davies state at early times. Keeping $\varepsilon$ nonzero also regulates the UV divergences appearing in the correlation functions, allowing us to keep track of their structure.
\item For applications to super-Hubble modes we then expand the result to obtain the asymptotic form near $z=0$. It is at this point that the Nakajima-Zwanzig becomes approximately Markovian at leading order.
\item One can choose $z_{\rm in}$ as one likes, even while choosing the Bunch Davies vacuum, but doing so with $z_{\rm in}$ chosen at some finite value corresponds to a `sudden' approximation wherein an initially adiabatic Bunch Davies state is very suddenly perturbed by rapidly switching on the interaction. We instead take the limit $z_{\rm in} \to \infty$ along the Maldacena contour since this is by far the most natural choice, choosing as it does the adiabatic evolution from the remote past. 
\item Finally, the last step is to expand about $\varepsilon=0$, since keeping nonzero $\varepsilon$ is important at intermediate steps for the reasons discussed in \cite{Burgess:2024heo}. It is once this expansion is made that UV divergences appear as powers or logarithms of $1/\varepsilon$.

\end{itemize}
 
\subsection{Correlation function results}

There are a number of correlation functions that need to be calculated. We display one such calculation here; the others follow the same pattern. Being explicit about a representative example allows us to show how the use of the deformed contour plays such an important role. The example we highlight here is $\scrC^{\ssD\ssD}_k$ obtained by Fourier transforming the environmental correlator $C^{\ssD\ssD}_k(\eta,\eta') = \langle B^\ssD(\bm{x},\eta) B^\ssD(\bm{x}^{\prime},\eta^{\prime})$ where $B^\ssD := \delta^{ij} \partial_{i} {\zeta}_{\env} \partial_{j} {\zeta}_{\env}$. This is also the operator considered in \cite{Burgess:2022nwu}, and so allows a comparison of the differences between that calculation and this one. The main difference is the restriction of $\eta_{\in}$ to be super-Hubble, and the care we must take here with the appropriate contour when the limit $\eta_{\in} \to - \infty$ is taken. The handling of divergences is also done differently here (see Appendix \ref{App:Comparison} for a detailed comparison).

Since we work in the interaction picture fields evolve using the free equations of motion and we can use Wick's theorem to write
\begin{equation}
\label{eq:DDWick}
C^{\ssD\ssD}(\bm{x},\eta;\bm{x}^{\prime},\eta^{\prime})=2 \Bigl\langle\partial_i {\zeta}_{\env}(\bm{x},\eta)\partial_j {\zeta}_{\env}(\bm{x}^{\prime},\eta^{\prime}) \Bigr\rangle \Bigl\langle\partial_i {\zeta}_{\env}(\bm{x},\eta)\partial_j {\zeta}_{\env}(\bm{x}^{\prime},\eta^{\prime}) \Bigr\rangle.
\end{equation}
Here the brackets denote vacuum expectation values taken in the Bunch-Davies state for the environmental part of the field. Next, we convert from $\zeta_\env$ to $v_\env$ and use the free field expansion of ${v}_{\env}(\bm{x},\eta)$ in terms of momentum modes and creation and annihilation operators, which imply: $\langle v_{\bm{k}}(\eta)\, v_{\bm{q}}(\eta^{\prime})\rangle = u_k(\eta)u^*_q(\eta^{\prime} )\delta_{\bm{k},-\bm{q}}$. Performing the Fourier transform one finds (after taking the continuum limit):
\begin{equation}
\label{eq:DDCorrKspace}
\scrC_k^{\ssD\ssD}(\eta,\eta^{\prime}) = 2\int \frac{\exd^3 q}{(2\pi)^3}\ \frac{\exd^3 l}{(2\pi)^3}\ \Bigl[  u_q(\eta) u^*_q(\eta^{\prime})\Bigr]\Bigl[   u_l(\eta) u^*_l(\eta^{\prime})\Bigr] (\bmq \cdot \bml)^2 (2\pi)^3 \delta^{3}\left(\bm{k}- \bm{q}-\bm{l}\right) \,.
\end{equation}

Writing the momentum delta function in Fourier space we arrive at a factorized form:
\begin{eqnarray}
\label{eq:DDCorrKspace}
\scrC_k^{\ssD\ssD}(\eta,\eta^{\prime})&=& 2 \int \exd^3 y\ e^{i \bm{k}\cdot\bm{y}}\ F_{i j}(\bm{y}) F_{i j}(\bm{y})\nonumber\\
\hbox{where} \qquad F_{i j}(\bm{y})&=&\int \frac{\exd^3 q}{(2\pi)^3}\ e^{-i \bm{q}\cdot\bm{y}}\; q_i q_j u_q(\eta) u^*_q(\eta^{\prime}) \,.
\end{eqnarray}
$F_{i j}(\bm{y})$ is a three-tensor depending only on $\bm{y}$ and so can be decomposed in terms of the (mutually orthogonal) projectors:
\be 
\label{eq:projectors}
\left(i j\right)  =  \delta_{i j}-\hat{y}_i \hat{y}_j \qquad \hbox{and} \qquad
\left[ij\right]  =  \hat{y}_i \hat{y}_j.
\ee
We find in this way
\be 
\label{eq:Fdecomp}
F_{i j}(\bm{y}) =  a(y) \left(i j\right)+b(y) \left[ij\right] \,,
\ee
with
\bea \label{abdefs}
2 a(y) &=& F_{i j}(\bm{y})\left(i j\right)=\frac{1}{(2\pi)^2} \int_{q>k_{\UV}} \exd q\ \int_{-1}^{+1} \exd\nu\ (1-\nu^2) e^{-i q y \nu} f_q(\eta,\eta^{\prime}) \\
b(y) &=& F_{i j}(\bm{y})\left[i j\right]=\frac{1}{(2\pi)^2} \int_{q>k_{\UV}} \exd q\ \int_{-1}^{+1} \exd\nu\ \nu^2 e^{-i q y \nu} f_q(\eta,\eta^{\prime})\nonumber
\eea
where
\be
f_q(\eta,\eta^{\prime}) :=  q^2 u_q(\eta) u^*_q(\eta^{\prime}) \qquad \hbox{and} \qquad
\nu :=  \hat{y}\cdot\hat{q}=\cos\theta_q \,.
\ee

The integral over the momentum $q$ converges because the factor $\exp[-i q (\eta-\eta^{\prime})]$, coming from the mode functions, damps the integrand exponentially due to the contour deformation $\eta^{\prime}\rightarrow \eta^{\prime}+i\varepsilon \left|\eta^{\prime}\right|$. Using this in \eqref{eq:DDCorrKspace} leads to
\begin{eqnarray}
\label{eq:contourexpressionDD}
\scrC_k^{\ssD\ssD}(\eta,\eta^{\prime})&=& -i \int_0^{\infty}\ \exd y\ y \left(e^{i k y}-e^{-i k y}\right)\Bigl[ 2 a^2(y)+2 b^2(y) \Bigr]\nonumber\\ 
&=& - i \int_{-\infty}^{\infty}\ \exd y\ y \left(e^{i k y}-e^{-i k y}\right)\Bigl[ a^2(y)+ b^2(y)\Bigr] \,,
\end{eqnarray}
where the last equality uses that the integrand is even under $y\leftrightarrow -y$. The remaining $y$ integral can be done by contour integration using the fact that $a^2+ b^2$ has poles at $y=0$ and $y=\pm \Delta\mp i \left|\eta^{\prime}\right|$, where $\Delta=\eta-\eta^{\prime}$. Furthermore, the constraint $k_{\UV}>k$ tells us how to close the contour of integration in the complex $y$ plane. 

A similar calculation can be done for all the correlation functions, which we have automated using Mathematica. The general result gives correlators $\scrC_k^{a b}(z,z')$ with the generic form described earlier -- {\it c.f.}~eqs.~\pref{eq:genericorrform} and \pref{eq:corrcoeffs} -- but repeated for convenience here:
\begin{equation}
\label{eq:genericorrformx}
\scrC_k^{a b}(z,z') = \frac{e^{-2 i \kappa w}}{8\pi^2} \Bigl[ \cC_0^{a b}(w,z)+e^{-i w} \cC_1^{a b}(w,z)\Bigr],
\end{equation}
where $\kappa=k_{\UV}\slash k$ and the coefficients $\cC_0^{a b}(w,z),\ \cC_1^{a b}(w,z)$ take the form:
\begin{equation}
\label{eq:corrcoeffsx}
\cC^{a b}_i = \sum_{u=-u_{\rm max}}^{u_{\rm min}} \frac{f^{ab}_{iu}(z)}{w^u} \,.
\end{equation}
This finally leads to the time integrals of the form 
\begin{equation}
\label{eq:genericintegralsx}
I_j =   \int_{-i\varepsilon z}^{z_{\in}-z-i\varepsilon z_{\in}} \frac{\exd w}{w^j} \; e^{-i \beta w}  \,,
\end{equation}
whose small-$z$ behaviour is discussed in \pref{eq:genericintegrals}.

\subsection{Calculation of $\mathcal{T}^{rs}_k(\eta,\eta')$ and $\cJ^{rs}_{kn}(\eta,\eta_{\in})$}

With these expressions for $\scrC^{ab}_k(z,z')$ the next step is to compute the combinations $\cT^{rs}_{k}(z,z')$ that appear in the evolution equation. To get $\cT^{rs}_k$ from $\scrC^{ab}_k$ we must group correlators according to which kernel -- $\cW_k$, $\cX_k$, $\cY_k$ or $\cZ_k$ of eq.~\pref{eq:etaprimeToeta2} -- that appears when evolving the fields, and exploit rotational invariance by decomposing the operators by their tensorial structure: rotational scalar ($j=0$), vector ($j=1$, with one factor of $\hat{\bmk}_i$) and tensor ($j=2$, with two factors $\hat{\bmk}_i \hat{\bmk}_j$). 

For instance, the terms in $\cT^{\zeta\zeta}_k$ involving the $\cW_k$ kernel contribution in eq.~\pref{eq:zetazeta} involve the correlators:
\begin{equation}
\label{eq:zetazetaAkernel}
\left\langle \left[\cE^{\zeta}+\cE^{\zeta i}\left(-i\hat{\bmk}_i k\right)\right]_{\eta} \left[\cE^{\zeta}+\cE^{\zeta j}\left(i\hat{\bmk}_j k\right)\right]_{\eta^{\prime}}\right\rangle,
\end{equation}
while the $\cZ_k$ kernel contributions arise in the combination
\begin{equation}
\label{eq:zetazetaDkernel}
\left\langle \left[\cE^{\zeta}+\cE^{\zeta i}\left(-i\hat{\bmk}_i k\right)\right]_{\eta} \left[\cE^{\mfp}+\cE^{\mfp j}\left(-i\frac{\hat{\bmk}_j}{ k}\right)\right]_{\eta^{\prime}}\right\rangle.
\end{equation}
From this we read off that the $j=0$, $\cW$-kernel contribution to $\cT^{\zeta\zeta}_k$ involves only the correlator
\begin{equation}
\label{eq:zetazetaAjzero}
\left\langle \cE^{\zeta}(\eta) \cE^{\zeta}(\eta^{\prime})\right\rangle \,,
\end{equation}
which we evaluate as described above. A similar evaluation gives the $j = 1$ and $j = 2$ contributions to $\cT^{\zeta\zeta}_{k}$ and the same logic also applies to the combinations $\cT^{\zeta\mfp}_{k}$,  $\cT^{\mfp\zeta}_{k}$ and $\cT^{\mfp\mfp}_{k}$. 

Once these are known we perform the integrations that give $\cJ^{rs}_{kn}(\eta,\eta_{\rm in})$. Given the argument made above that higher terms in the series $(\eta' - \eta)^n \, \partial_\eta^n \varrho$ are suppressed by additional powers of $z$ once the integral in $\cJ^{rs}_{kn}$ are performed we now specialize to the super-Hubble regime $z \ll 1$ and focus only on the dimensionless functions with $n=0$: $\widetilde\cJ^{rs}_{k0}(z)$. When doing so we find that the Laurent series in powers of $z$ for $\widetilde\cJ^{\zeta\zeta}_{k0}$ starts at order $z^{-4}$, while that for $\widetilde\cJ^{\zeta\mfp}_{k0}$ starts at $z^{-2}$, that for $\widetilde\cJ^{\mfp\zeta}_{k0}$ starts at $z^{-1}$ and $\widetilde\cJ^{\mfp\mfp}_{k0}$ starts at $z$. We focus here on describing the calculation of $\widetilde\cJ^{\zeta\zeta}_{k0}$ in detail, since this is the one that dominates for small $z$. 

Evaluating the correlator of \pref{eq:zetazetaAjzero} and performing the $w$ integrals in the small-$z$ limit we find the leading terms in the Laurent series of the $j=0$ contribution for small $z$ are given by 
\begin{subequations} \label{eq:zetazetaAjsSHL0}
\begin{align}
{\rm Re}\left(\widetilde\cJ^{\zeta\zeta}_{k0}\right)_{j=0,\cW} &=\frac{8}{\varepsilon z^4} +\frac{1}{z^3} \left(-6 -8\kappa-\frac{14}{3\kappa} \right)+\frac{1}{z^2}\left(\frac{4\pi}{3}\right)- \frac{2\pi}{15}+ \cdots \,,\nn\\
{\rm Im}\left(\widetilde\cJ^{\zeta\zeta}_{k0}\right)_{j=0,\cW} &=\frac{8-\frac{4}{\varepsilon^2}}{z^4}+\frac{4\pi}{3 z^3}+ \frac{1}{z^2}\left[ \frac{13}{3}-\frac{4}{\kappa}+12\kappa +8\kappa^2-\frac{16}{3} \Bigl( \ln(2\kappa z)+\gamma_\ssE+\ln\varepsilon \Bigr)\right] \nn\\
&\qquad -\frac{2\pi}{15 z}+\left( \frac{2}{5 \kappa}+\frac{8}{15}\Bigl(\ln\varepsilon+\ln(2\kappa z)+\gamma_\ssE \Bigr)\right) +\cdots \,.
\end{align}
\end{subequations}
Repeating the exercise for the $j=1$ contributions similarly gives
\begin{subequations} \label{eq:zetazetaAjsSHL1}
\begin{align}
{\rm Re}\left(\widetilde\cJ^{\zeta\zeta}_{k0}\right)_{j=1,\cW} &=\frac{-\frac{32}{\kappa}+\frac{8}{\kappa}\ln(\frac{2}{\delta p})}{ z^3} +\frac{-\frac{14\pi}{3}+12\pi\ln(\frac{2}{\delta p})}{z^2}-\frac{8}{3\kappa z}-\frac{16\pi}{15} \cdots \,,\\
{\rm Im}\left(\widetilde\cJ^{\zeta\zeta}_{k0}\right)_{j=1,\cW} &=\frac{6\pi-4\pi \ln(\frac{2}{\delta p})}{z^3}+\frac{1}{z^2}\left(\frac{8}{3}-\frac{31}{\kappa}-\frac{8}{\kappa}\ln(2 \delta p)-\frac{88}{3}\Bigl( \ln(2\kappa z)+\gamma_\ssE+\ln\varepsilon \Bigr) \right)\\
&\qquad +\frac{4\pi}{3 z}+\frac{7}{15\kappa}+\frac{16}{15}\Bigl( \ln(2\kappa z)+\gamma_\ssE+\ln\varepsilon \Bigr) \cdots \,,
\end{align}
\end{subequations}
while the $j=2$ expression is
\begin{subequations} \label{eq:zetazetaAjsSHL2}
\begin{align}
{\rm Re}\left(\widetilde\cJ^{\zeta\zeta}_{k0}\right)_{j=2,\cW} &= -\left(\frac{8}{\kappa z^3}\right)\left(1-\frac{2}{\delta p}+2\ln(\frac{2}{\delta p})\right)+ \frac{16\pi}{\delta p\  z^2}-\frac{8}{3 \kappa z} -\frac{16\pi}{5}+\cdots \,,\\
{\rm Im}\left(\widetilde\cJ^{\zeta\zeta}_{k0}\right)_{j=2,\cW} &=\frac{4\pi-\frac{8\pi}{\delta p}}{ z^3}-\frac{8}{\kappa z^2} \Big( 1+\frac{1}{\delta p}+\ln(2\kappa \delta p) \Big) +\frac{16}{5}\Bigl(\frac{1}{3 \kappa} + \ln(2\kappa z \varepsilon) +\gamma_\ssE \Bigr)+\cdots
\end{align}
\end{subequations}
where $\gamma_\ssE$ denotes the Euler-Mascheroni constant.

Notice the appearance of terms that are singular in the limit $\varepsilon \to 0$, reflecting divergences in the integrations brought on by the singular dependence of correlators in the coincidence limit. There is also a dependence on momentum regulators $\delta p,\ \delta_0$, where $\delta p,\ \delta_0\rightarrow 0$ signals a kinematic divergence occurring when the environmental momenta are back to back. We expect that when all terms of the appropriate order in $\slrl$ are included in an observable, both the $\varepsilon$ as well as the $\delta p$ dependences will all cancel. However verifying this requires the calculation of the contributions from scalar-tensor interactions as well as higher order corrections to the modes.

Summing these contributions -- and repeating the exercise for the other correlators -- gives the desired expressions for the small-$z$ limit once all of these contributing pieces are summed (all these expressions are multiplied by a factor of $2\slash\pi^2$ coming from normalizations).  
\begin{subequations} \label{FinalSmallzzetazeta}
\begin{align}
{\rm Re}\left(\widetilde\cJ^{\zeta\zeta}_{k0}\right) &= \frac{8}{\varepsilon z^4} +\frac{-6+\frac{16}{\kappa}-8\kappa-\frac{24}{\delta_0}}{z^3}+\frac{-\frac{2\pi}{3}-\frac{12\pi}{\delta_0}+\frac{16\pi}{\delta p}+\frac{40}{3\varepsilon}+\frac{8\pi}{\kappa}+16\pi\ln\frac{2}{\delta p}}{z^2}\nn\\
& +\frac{-10-\frac{136}{15\kappa}-\frac{40\kappa}{3}}{z}-\frac{76\pi}{15}\\
{\rm Im}\left(\widetilde\cJ^{\zeta\zeta}_{k0}\right) &=\frac{8-\frac{4}{\varepsilon^2}}{z^4}+\frac{2\pi}{z^3}+\frac{-21-\frac{83}{2\kappa}+12\kappa+8\kappa^2+\frac{16}{\kappa\ \delta p}+\frac{12}{\kappa}\ln \delta_0-\frac{40}{3}\left(\ln\left(2\kappa z\right)+\ln\varepsilon+\gamma_\ssE \right)}{z^2}\nn\\
&-\frac{8\pi}{3 z}+\frac{104}{15}-\frac{74}{15\kappa}-\frac{32}{15}\left( \ln\left(2\kappa z\right)+\ln\varepsilon+\gamma_\ssE \right)+\cdots \,.  
\end{align}
\end{subequations}
The corresponding results for the other correlators are obtained in a similar way and give the following small-$z$ expansion:
\begin{subequations} \label{FinalSmallzzetap}
\begin{align}
{\rm Re}\left(\widetilde\cJ^{\zeta\mfp}_{k0}\right) &=\frac{\frac{224}{3\kappa}-\frac{60}{\kappa\ \delta_0}}{z^2}+\frac{128}{3 \varepsilon z}-38-\frac{904}{15\kappa}-\frac{152\kappa}{3}+\cdots, \\
{\rm Im}\left(\widetilde\cJ^{\zeta\mfp}_{k0}\right) &=\frac{2\pi}{z^2}+\frac{-\frac{64}{3}+\frac{20}{3\varepsilon^2}}{z}-\frac{2\pi}{3}+\cdots \,,
\end{align}
\end{subequations}
\begin{subequations} \label{FinalSmallzpzeta}
\begin{align}
{\rm Re}\left(\widetilde\cJ^{\mfp\zeta}_{k0}\right) &=-\frac{2}{3 \varepsilon z}+\frac{11}{16}-\frac{19}{30\kappa}+\frac{1}{2\delta_0}+\frac{11\kappa}{12}+\cdots\\
{\rm Im}\left(\widetilde\cJ^{\mfp\zeta}_{k0}\right) &=\frac{-\frac{2}{3}+\frac{5}{24\kappa^2}}{z}-\frac{5\pi}{24}+\cdots 
\end{align}
\end{subequations}
\begin{subequations} \label{FinalSmallzpp}
\begin{align}
{\rm Re}\left(\widetilde\cJ^{\mfp\mfp}_{k0}\right) &= z \left(\frac{1709}{1280\kappa }+\frac{\pi ^2}{128}\right)+\cdots\\
{\rm Im}\left(\widetilde\cJ^{\mfp\mfp}_{k0}\right) &=z\left( \frac{3 \pi }{128 \kappa }+\frac{\pi}{64}  \ln (2 \kappa)-\frac{379 \pi}{768}\right)+\cdots \,.
\end{align}
\end{subequations}
These show that the leading contribution for the real and imaginary part of each correlator is UV divergent and dependent of $k_\UV$. The correlator that grows the fastest as $z \to 0$ is the real part of the $\zeta$-$\zeta$ correlator, which scales as $z^{-4}$.

We expect that some terms in the full result do diverge as $\varepsilon\rightarrow 0$, but the divergences only show up in terms that are subdominant in $z$ and so compete with other subdominant contributions like the non-Markovian terms involving $\partial_\eta^n \varrho$ (which are also divergent). All UV divergences must in the end be absorbed into renormalizations of couplings in $\scrH_{\rm eff}$. Since hamiltonian evolution cannot contribute to the decoherence this ensures that the leading contribution to the purity does not compete with any counterterms and so must necessarily be positive and UV finite within a reliable approximation scheme, though proving this requires also computing the contributions from environmental tensor modes not computed here.  

\section{Implications for the purity}
\label{sec:PurityCalc}

The final step is to compute the implications of the evolution equation \pref{eq:lindbladlikeform} for the purity of the system's state, as given in \pref{puritydef}. This discussion goes through here much as it did in \cite{Burgess:2022nwu}, so we mostly highlight the changes. 

We wish to compute how the purity \pref{puritydef} evolves in time, starting from an initially pure Bunch-Davies state in the remote past. We take advantage of the intrinsic gaussianity of the evolution -- as described in \S\ref{sssec:Gaussianity} -- to be able to do so mode by mode. To that end we differentiate \pref{purityk} and use \pref{eq:lindbladlikeform} to evaluate the result, leading to
\be \label{LindbladPurity}
  \partial_\eta \gamma_\bmk := 2 \hbox{Tr}_{\rm sys} \left( \varrho_\bmk \, \partial_\eta \varrho_\bmk \right) 
  \simeq 2 \sum_{r,s=1}^2 h^{s r}_\bmk\left[ \hbox{Tr}_{\rm sys} \Bigl(  \varrho_{\bmk} \, \mathcal{O}^{\dag}_{\bmk,r}\varrho_{\bmk} \,  \mathcal{O}_{\bmk,s} \Bigr) - \hbox{Tr}_{\rm sys} \Bigl( \varrho_\bmk^2 \, \mathcal{O}^{\dag}_{\bmk,r}\mathcal{O}_{\bmk,s} \Bigr) \right]\,.
\ee
The commutator term from the right-hand side of \pref{eq:lindbladlikeform} drops out because of the cyclic property of the trace.

\subsection{Perturbative evolution}

Evaluating the right-hand side of \pref{LindbladPurity} is simplest when working with straight-up perturbation theory because in this case the appearance of the small explicit factor of $\slrl (H/\Mp)^2$ within $h_\bmk^{rs}$ -- {\it c.f.}~eq.~\pref{eq:decohmatrixh} -- means it suffices on the right-hand side to use the lowest-order expression: the initial value for the system state: $\varrho_{\bmk0} = | \Omega \rangle \, \langle  \Omega |$. This is a pure state that we choose to be the Bunch-Davies state, $|\Omega \rangle$, satisfying $c_\bmk | \Omega \rangle = 0$. 

Doing so simplifies both terms in \pref{LindbladPurity}. It simplifies the first term on the right-hand side because it allows it to be written 
\be
   \hbox{Tr}_{\rm sys} \Bigl(  \varrho_{\bmk} \, \mathcal{O}^{\dag}_{\bmk,r}\varrho_{\bmk} \,  \mathcal{O}_{\bmk,s} \Bigr) = \langle  \, \mathcal{O}^{\dag}_{\bmk,r}  \rangle \, \langle   \mathcal{O}_{\bmk,s} \rangle \,,
\ee
where $\langle \cdots \rangle := \langle \Omega | \cdots | \Omega \rangle$. This expectation value vanishes in the Bunch-Davies state for both $\cO_1 \propto \zeta$ and $\cO_2 \propto \mfp$. Starting with a pure state also simplifies the second term on the right-hand side of \pref{LindbladPurity} because it implies $\varrho_\bmk^2 = \varrho_\bmk$, allowing the second terms to be interpreted as an expectation value. 

We find in this way the central result
\be  \label{LindbladPurity2}
  \partial_\eta \gamma_\bmk   \simeq  - 2 \sum_{r,s=1}^2 h^{s r}_\bmk  \; \Bigl\langle  \mathcal{O}^{\dag}_{\bmk,r}\mathcal{O}_{\bmk,s} \Bigr\rangle  \,,
\ee
and so evaluating the coefficients $h^{s r}$ using \pref{eq:decohmatrixh} implies $\partial_z \gamma_\bmk =- \partial_\eta \gamma_\bmk/k$ evaluates to
\bea \label{LindbladPurity3}
\partial_z \gamma_\bmk &\simeq& 
\frac{\epsilon_1 H^2}{2M_p^2}  \bigg[ \tfrac12 \mathrm{Im}[\widetilde{\mathcal{J}}^{\mfp\zeta}_{k0}  - \widetilde{\mathcal{J}}^{\zeta\mfp}_{k0}  ]  + \mathrm{Re}[\widetilde{\mathcal{J}}^{\zeta \zeta}_{k0}] \,  Z^{-2} \Bigl\langle \zeta_{\bmk} \zeta_{-\bmk} \Bigr\rangle\\
&& \qquad\qquad\qquad  +  \mathrm{Re}[\widetilde{\mathcal{J}}^{\zeta \mfp}_{k0}+\widetilde{\mathcal{J}}^{\mfp \zeta}_{k0}]  \, \left\langle  \tfrac12 \Bigl\{\zeta_{\bmk}, \mfp_{-\bmk} \Bigr\} \right\rangle  + \mathrm{Re}[\widetilde{\mathcal{J}}^{\mfp \mfp}_{k0}] \,  Z^{2} \,\Bigl\langle \mfp_{\bmk} \mfp_{-\bmk} \Bigr\rangle  \bigg] \nn
\eea
where all operators are evaluated at $\eta$ (or $z = - k \eta$) and  $Z^2 =H^2/(2 \epsilon_1 k^3 M_p^2)$, as defined in \pref{OiandZdefs}.
 
The leading system correlators can be evaluated using the mode functions given in \pref{eq:deSitter:BD:vk},  
\be
  \Bigl\langle  \zeta_{-\bmk} \zeta_{\bmk} \Bigr\rangle = | \hat u_k |^2 
   = Z^2(1+z^2) \,, \qquad \Bigl\langle  \mfp_{-\bmk} \mfp_{\bmk} \Bigr\rangle = \mfz^4 | \hat u'_k |^2 
   = \frac{ Z^{-2}}{z^2} \,,   
\ee
and
\be
  \Bigl\langle  \zeta_{-\bmk} \mfp_{\bmk} \Bigr\rangle = \mfz^2 \hat u_k \, (\hat u'_k)^* = \frac{1}{z}\,(1+iz) = \Bigl\langle  \mfp_{-\bmk} \zeta_{\bmk} \Bigr\rangle^* \quad \hbox{so} \quad  \Bigl\langle \tfrac12\{ \zeta_{-\bmk} , \mfp_\bmk \} \Bigr\rangle = \frac{1}{z} \,.   
\ee
and using these in \pref{LindbladPurity3} finally gives
\be \label{LindbladPurity4}
\partial_z \gamma_\bmk  \simeq  
\frac{\epsilon_1 H^2}{2M_p^2}  \left[  (1+z^2) \mathrm{Re}[\widetilde{\mathcal{J}}^{\zeta \zeta}_{k0}]  + \frac{1}{z} \, \mathrm{Re}[\widetilde{\mathcal{J}}^{\zeta \mfp}_{k0}+\widetilde{\mathcal{J}}^{\mfp \zeta}_{k0}]   + \frac{1}{z^2} \mathrm{Re}[\widetilde{\mathcal{J}}^{\mfp \mfp}_{k 0}]   + \tfrac12 \mathrm{Im}[\widetilde{\mathcal{J}}^{\mfp\zeta}_{k0}  - \widetilde{\mathcal{J}}^{\zeta\mfp}_{k0}  ]   \right] \,. 
\ee
Inserting the explicit small-$z$ expressions \pref{FinalSmallzzetazeta} through \pref{FinalSmallzpp} for the small-$z$ limit of the $\widetilde \cJ^{rs}_{k0}$ coefficients shows that the only term in this equation that dominates is the one involving the real part of $\widetilde \cJ^{\zeta\zeta}_{k0}$, and this contributes a contribution that is order $z^{-4}$. All of the other terms are down by at least a factor of $z$. The total coefficient of $z^{-4}$ read off from \pref{FinalSmallzzetazeta} is $8/\varepsilon$, leading to the super-Hubble expression  
\be   \label{LindbladPurity5}
\partial_z \gamma_\bmk  =  \frac{4 \slrl H^2}{\varepsilon z^{4}\Mp^2} \Bigl[ 1 + \cO(z) \Bigr] \qquad  \hbox{when} \; z = - k\eta = \frac{k}{aH} \ll 1   \,.
\ee  
To this must be added the contribution coming from scalar interactions with environmental tensor modes, whose addition is expected to cancel the divergences, leaving an undetermined finite remainder.  

\section{Lessons learned}
\label{sec:Conclusions}

In this paper we report the result of a calculation of the leading decoherence of long-wavelength scalar-metric fluctuations due to their gravitational interactions with short wavelength scalar-metric perturbations. We do so within a minimal inflationary framework and because we allow system and environmental modes to be both super-Hubble and sub-Hubble we are able to explicitly track system-environment interactions into the remote past to make contact with the Bunch-Davies initial conditions. For applications to primordial fluctuations we specialize to the late-time regime when the decohering long-wavelength modes are super-Hubble, for which we find several simplifications.

What is new in this calculation is the inclusion of {\it all} interactions that can contribute to lowest order in the semiclassical expansion. This allows us to identify systematically which interactions decohere most efficiently, which parts of the environment are responsible and when the decoherence occurs. Our calculation is informative about the behaviour of primordial quantum fluctuations within inflationary models and also for open-quantum-system calculations more generally. The following sections briefly summarize both types of insights and highlight open directions for future work.

\subsection{For inflationary primordial fluctuations}
\label{ssec:LLInf}

Our calculation reveals several characteristic properties of gravitation-mediated decoherence within minimal inflationary models:
\begin{itemize}
\item Semiclassical perturbation theory organizes the calculation into powers of $H^2/M_p^2$, where $H$ is the inflationary Hubble scale. For GR coupled to a scalar field -- as found in minimal inflationary models -- the leading nontrivial contribution arises at order $H^2/M_p^2$ and is mediated by interactions cubic in the fields. Choosing the environment to be short-wavelength modes then ensures the relevant interactions are strictly linear in the system-variable fields and this in turn implies the effective evolution of the purity of long-wavelength system modes at this order is gaussian. This in turn implies that it can be understood mode-by-mode without mode-mixing.
\item Eq.~\pref{LindbladPurity5}, reproduced here as
\be   \label{LindbladPurityRecap}
\partial_z \gamma_\bmk   =   \frac{4 \slrl H^2}{\varepsilon z^{4}\Mp^2} \Bigl[ 1 + \cO(z) \Bigr] \qquad  \hbox{when} \; z = - k\eta = \frac{k}{aH} \ll 1   \,.
\ee  
is our main perturbative result for the evolution of the purity of long-wavelength modes. It is universal in the sense that it does not depend on the details of the split between system and environment (like the value of $k_\UV$). But it also diverges due to the presence of the factor $1/\varepsilon$, indicating that it must compete with other divergent contributions that arise when other interactions (like the coupling to a tensor environment) are included. As discussed in the main text, obtaining this expression requires careful handling of the initial conditions to properly project onto the Bunch Davies vacuum at early times.
\item Eq.~\pref{LindbladPurityRecap} shows that decoherence for super-Hubble modes during inflation grows very quickly with time and so is ruthlessly efficient despite being mediated only by very weak gravitational interactions. Although the prefactor $\epsilon_1 H^2/M_p^2$ is at most of order $10^{-14}$ the growth like $a^3 \propto e^{3Ht}$ is fast enough that perturbation theory fails after only a handful of $e$-foldings.
\item The $a^3$ growth resembles the $a^3$ behaviour found in earlier partial calculations like the ones in \cite{Nelson:2016kjm, Burgess:2022nwu}, though this need not remain so once the uncomputed contributions from the tensor environment are included. The differences between this calculation and the one in \cite{Burgess:2022nwu} is described in detail in Appendix \ref{App:Comparison}, but main reason is because the dominant interaction responsible for \pref{LindbladPurityRecap} turns out to be the `nonlocal' interactions obtained by using the constraints to eliminate the lapse and the shift variables, whereas earlier calculations specialized to a specific local interaction. The important role played by the constraints contains an echo of the closely related calculations in \cite{Danielson:2021egj, Danielson:2022tdw, Danielson:2022sga} of decoherence induced by the presence of horizons.
\item The leading behaviour for super-Hubble modes is Markovian, with deviations from the Markovian limit controlled by powers of $k/(aH)$. The leading deviations already arise at first subleading order -- they are suppressed only by a single power of $k/(aH)$. Non-Markovian effects during inflation are present but suppressed for super-Hubble modes by powers of $k/(aH)$ (though not by many of those powers).
\end{itemize}

\subsection{More generally}
\label{ssec:MoreGeneral}

Our calculation also illustrates several properties of open-system effects in general -- and of decoherence calculations in particular -- that apply more widely than just to inflationary applications. 
\begin{itemize}
\item Similar to \cite{Burgess:2022nwu} -- but unlike, say, \cite{Nelson:2016kjm} -- we expect the leading contributions to decoherence to be UV finite, despite UV divergences arising in many intermediate steps. The appearance of divergences here suggest that the uncomputed tensor parts cannot be neglected.\footnote{This expectation is borne out by the dominant tensor contributions, which do prove to be UV finite \cite{ToAppear}.} This is as expected since the decoherence calculation arises at loop order in the general semiclassical $H^2/M_p^2$ expansion \cite{Burgess:2003jk, Burgess:2009ea, Adshead:2017srh}. UV finiteness is required at leading order because general arguments ensure that UV divergences can be absorbed into the renormalization of counterterms in the effective lagrangian, but decoherence can never arise due to a particular choice for a value of a coupling in the lagrangian. This does {\it not} mean that UV divergences never arise at all, however. Indeed, the purity evolution necessarily depends explicitly on couplings like $M_p$ and so must eventually diverge just because these couplings themselves receive divergent renormalizations. But these divergences first arise at subdominant order in the semiclassical expansion, where they cancel explicit divergences found in the higher-order graphs required at subdominant order. 
\item The discussion of divergences also highlights another subtlety of decoherence calculations within field theory: their sensitivity to nonlinear field redefinitions. It is very easy when computing the evolution of the purity of long-wavelength states due to a short-wavelength environment to find spurious results where decoherence appears to be much stronger than (or weaker than) it really is  \cite{Burgess:2024eng}. The problem is that naively integrating out a heavy field involves making very specific choices of how high-energy modes are treated because interactions inevitably mix up low- and high-energy basis states. One must be careful when doing so to choose one's ordering of limits so that spurious high-energy ambiguities decouple in the way that they are required to on general grounds \cite{Burgess:2024heo}. Part of this choice requires a robust projection onto the adiabatic vacuum -- the Bunch Davies state in the inflationary case -- as we are careful to do in the main text.
\item A special case of the above discussion is the source of some confusion in the literature. When categorizing interactions -- such as the discussion of cubic interactions within inflation in \cite{Maldacena:2002vr} -- it is common to drop total time derivatives in the action. This can be confusing because total derivatives are also known to generate nonzero contributions to the purity \cite{Sou:2022nsd}, making their neglect seem suspicious. Neglecting total time derivatives is ultimately justified because they can be removed by performing an appropriate canonical transformation of the underlying field theory and physical predictions are unaffected by canonical transformations (as has been verified in detail in cosmology \cite{Braglia:2024zsl}). But although the purity of a system is unchanged by canonical transformations that map system and environment into themselves, it can change when the transformation maps system into environment and vice versa. Sensitivity of decoherence to total derivatives is a special case of the observation of the previous bullet point: decoherence calculations can also be sensitive to field redefinitions (and other UV details) if one is not careful to order the calculation in such a way as to systematically project onto the adiabatic ground state.
\item Finally, we believe the calculations presented here provide the tools required to settle the conceptual issue of the domain of validity of the Markovian approximation for the evolution of super-Hubble modes, at least insofar as decoherence calculations are concerned. Markovianity is not an assumption -- it is a consequence of the approximations that enter into the super-Hubble limit. When these approximations lead to a Lindblad equation this equation has only positive eigenvalues because unitarity is a property of the full theory before the approximations were made. There is no need in particular for appeals to other arguments like the `rotating wave approximation' that often arise when open-system tools are used in other areas of physics.\footnote{In particular all contributions to the purity evolution must be included when computing this expansion; it is not sufficient to keep only part of it (as was effectively done in \cite{Burgess:2022nwu} and recently discussed in \cite{Lopez:2025arw}).}
\end{itemize}

\subsection{Open issues}
\label{ssec:OpenIssues}

The story of inflationary decoherence does not end here, since this calculation leaves many issues untouched, most notably the explicit resummation of the purity to late times, as is required to predict its values at the end of inflation. 

We also do not ask here how short-wavelength tensor modes decohere long-wavelength scalar-metric modes. This issue was addressed in \cite{Burgess:2022nwu} for a specific type of cubic interaction, where each tensor mode was found to contribute the same as the scalar modes and so their addition simply amounted to multiplying the scalar result by 3\footnote{
Note added: The full calculation proves to go beyond what was found in \cite{Burgess:2022nwu} and to be more interesting in the sense that it contains terms that are less slow-roll suppressed than what is found here \cite{ToAppear}.}.

Our calculation here leaves open the size of decoherence for long-wavelength tensor modes, both from short-wavelength tensor and scalar environments. Again \cite{Burgess:2022nwu} took a step towards remedying this by computing how a scalar environment decoheres tensors through a specific type of interaction, leading to a result like \pref{LindbladPurityRecap} but unsuppressed by the slow-roll parameter $\epsilon_1$. Although this lack of suppression is likely to remain true (because $\epsilon_1$ also does not appear in the other possible interactions couplings these modes) a calculation along the lines given here is required to pin down the correct numerical coefficient. Work along these lines to determine this coefficient is ongoing.

A further question asks how the purity evolves in the post-inflationary environment before eventually re-entering the horizon. This is not captured by the calculations described above because these rely at various points on the slow-roll approximation that $\epsilon_1$ is small. This is not required conceptually however and so our calculation could be repeated to ascertain whether the decoherence we find survives subsequent post-inflationary evolution. 

We consider it a worthwhile exercise to pin down these issues, even if they only imply that states on horizon re-entry are strongly decohered and so quantum effects are unmeasurable in the later universe. In the end what is useful is to know how decoherence depends on the parameters of any underlying model, since this is what is required to determine what will be learned by any future attempts to measure it. 

\section*{Acknowledgements}
We thank Sebastian Cespedes, Thomas Colas, and Arnab Rudra for helpful conversations, and Mehrdad Mirbabayi and Enrico Pajer for insisting on the inclusion of time-dependent operators in our earlier decoherence calculations. CB's research was partially supported by funds from the Natural Sciences and Engineering Research Council (NSERC) of Canada. Research at the Perimeter Institute is supported in part by the Government of Canada through NSERC and by the Province of Ontario through MRI. G.K.~is supported by the Simons Foundation award ID 555326 under the Simons Foundation Origins of the Universe initiative, Cosmology Beyond Einstein's Theory as well as by the European Union Horizon 2020 Research Council grant 724659 MassiveCosmo ERC2016COG.

\appendix
%
\addtocontents{toc}{\setcounter{tocdepth}{1}}

\section{Cubic interactions amongst fluctuations}
\label{App:operators}

This appendix sketches the derivation of the standard EFT of scalar and tensor fluctuations for single-field
inflation, culminating in the list of cubic interactions given in \cite{Maldacena:2002vr}.  

\subsection{Single-field inflation}

As described in \S\ref{sec:OpenEFT}, single-field inflation consists of gravity and a single scalar field in the form of the action (\ref{actionstart}), repeated here for convenience,
\begin{equation}
\label{actionAppstart}
S[g_{\mu \nu},\varphi] = \int \exd^4 x\; \sqrt{ - g}
\biggl[ \frac{\Mp^2}{2} R - \frac{1}{2}
g^{\mu\nu} \, \partial_{\mu} \varphi \, \partial_{\nu} \varphi
- V(\varphi) \biggr] \ .
\end{equation}
Homogeneous classical solutions for the inflaton $\phi(t)$ and Hubble parameter $H = \dot{a}/a$ therefore obey
\begin{equation}
\label{background_App}
3 \Mp^2 H^2  =  \tfrac{1}{2} \dot{\phi}^2 + V(\phi) \,, \quad
\Mp^2 \dot{H}  =  - \tfrac{1}{2} \dot{\phi}^2  \quad \hbox{and} \quad
\ddot{\phi} + 3 H \dot{\phi} + V^{\prime}(\phi) = 0 \,,
\end{equation}
where dots mean derivatives with respect to cosmic time.

We perturb about a near-de Sitter spacetime, $\exd  s^2 = - \exd  t^2 +
a^2(t) \exd  \bm{x}^2$, working in the Arnowitt-Deser-Misner (ADM)
formalism as in \cite{Maldacena:2002vr} using the perturbed metric
\begin{equation}
  \exd  s^2  = - N^2 \exd  t ^2+ h_{ij} \left( N^i \exd  t + \exd  x^i \right)
  \left( N^j \exd  t + \exd  x^j \right) \ , 
\end{equation}
with $N$ the lapse function and $N^{i}$ the shift vector and the
inverse of spatial metric $h^{ij}$ defined by $h^{ij}h_{jk} =
\delta^{i}_{\; k}$. In terms of these variables the action
(\ref{actionAppstart}) becomes
\begin{align}
\label{actionApp}
S & = \int \exd^4 x\; \sqrt{h} N \left[\frac{\Mp^2}{2}
  \left( \cR  + \frac{E_{ij}E^{ij} - E^2}{N^2} \right) 
  + \frac{\big( \dot{\varphi} - N^{i} \partial_{i} \varphi\big)^2}{2 N^2}
  - \frac{1}{2} h^{ij} \partial_{i} \varphi \, \partial_{j} \varphi - V(\varphi)\right]\, 
\end{align}
with $\cR$ the 3D Ricci scalar built from the spatial metric $h_{ij}$
and $K_{ij} = E_{ij} / N$ is the extrinsic curvature of these spatial
slices, where
\begin{equation}
  E_{ij} := \sfrac{1}{2} \left(\dot{h}_{ij} - \nabla_{i} N_{j}
  -  \nabla_{j} N_{i}  \right) \quad \hbox{and} \quad E:= h^{ij} E_{ij} \ .
\end{equation}

Specializing to the gauge where the inflaton has no perturbation,
$\delta \varphi = 0$ and so $\varphi = \phi(t)$, the vanishing spatial
derivative $\partial_{j}\varphi =0$ allows the action to be simplified
to
\begin{align}
\label{actionApp2}
S & = \int \exd^4 x\; \sqrt{h} N
\left[ \frac{\Mp^2}{2} \left( \cR
  + \frac{E_{ij}E^{ij} - E^2}{N^2} \right) - V(\phi)
  + \frac{\dot{\phi}^2}{2 N^2} \right] \ .
\end{align}
The constraint equations obtained by varying $N$ and $N^{i}$ then become
\begin{equation}
\label{constraints}
\nabla_{i} \left( \frac{E^{i}_{\; j} - \delta^{i}_{\; j} E}{N} \right)
= 0, \quad \frac{\Mp^2}{2} \left( \cR + \frac{E_{ij}E^{ij} - E^2}{N^2}
\right) - V(\phi) + \frac{ \dot{\phi}^2}{2N^2} = 0 \,.
\end{equation}

These constraint equations are solved for $N$ and $N^{i}$ as functions
of the physical variables $\zeta$ and $\gamma_{ij}$, defined by
\begin{equation} \label{generalgauge}
  h_{ij} = a^2 e^{2 \zeta} \left( \delta_{ij} + \gamma_{ij}
  + \frac{1}{2} \gamma_{i\ell}\gamma_{\ell j} + \ldots \right) \,,
\end{equation}
with $\partial_{i} \gamma_{ij} = \gamma_{ii} = 0$. The goal is to
express the action~\eqref{actionApp2} as a function of these variables
after eliminating $N$ and $N^i$ using the constraints. Since our focus
is mainly on scalar fluctuations we drop $\gamma_{ij}$ in what
follows, simply quoting when needed the graviton-dependent terms found
elsewhere~\cite{Maldacena:2002vr}. For the metric (\ref{generalgauge})
the following relations prove useful:
\begin{equation}
\sqrt{h} = a^3 e^{3\zeta}, \quad   \cR   = a^{-2} e^{ - 2 \zeta}
\left[ - 4 ( \partial^2 \zeta ) - 2 ( \partial \zeta )^2 \right] \,,
\end{equation}
where ``$\partial$'' here denotes spatial differentiation.

\subsection{Quadratic scalar action}

We first verify the standard quadratic action for $\zeta$. At leading
order in $\zeta$ the lapse and shift are
\begin{equation}
\label{lapseshiftanswer}
N \simeq 1 + \frac{\dot{\zeta}}{H}, \quad
N_{i} \simeq -  \frac{\partial_{i} \zeta}{a^{2}H} + \partial_i \chi \,,
\end{equation}
where the field $\chi$ is defined as a solution to the equation
$\partial^2 \chi = \dot{\phi}^2 \dot{\zeta}/(2H^2\Mp^2)$, and so
\begin{equation}
\label{chidef}
\chi := \frac{\dot{\phi}^2}{2H^2\Mp^2} \partial^{-2}
\dot{\zeta} = \slrl  \partial^{-2} \dot{\zeta}  \ .
\end{equation}
Using the background equations of motion (\ref{background_App}) and
integrating by parts gives the quadratic action
\begin{equation}
\label{freescalaraction_App}
{}^{(2)}S = \int \exd  t \; \exd^3 \bm{x}\; \Mp^2 \slrl
\left[ a^3 \dot{\zeta}^2 - a (\partial \zeta )^2 \right]
\end{equation}
as given as \pref{freescalaraction} in the main text, with $\slrl = -
\dot H/H^2 = {\dot{\phi}^2}/({2H^2 \Mp^2})$. This becomes canonical after switching to conformal time and
once expressed in terms of the Mukhanov-Sasaki field
\begin{equation}
\label{MS_field_App}
v  = a \Mp \sqrt{2 \slrl } \zeta \,.
\end{equation}
When specialized to near-de Sitter geometries the quadratic action then is
\begin{equation}
{}^{(2)}S \simeq \int \exd  \eta \; \exd^3 \bm{x}\; \left[ \tfrac{1}{2} (v')^2
- \tfrac12 (\partial v)^2 + \frac{v^2}{\eta^2}  \right]  \ .
\end{equation}

\subsection{Cubic scalar interactions}
\label{ssec:AppCubic}

We next record the cubic self-interactions contained in the cubic part
of the expansion $S ={}^{(2)}S +{}^{(3)}S + \ldots$ of the
action in powers of the fluctuations found in
\cite{Maldacena:2002vr}. The part cubic in the scalar perturbation can
be written
\begin{align}
\label{allscalarcubics}
{}^{(3)}S & = \int \exd  t\, \exd^3 \bm{x}\, \Biggl\lbrace
\slrl ^2 \Mp^2 \left[ a \left(\partial \zeta\right)^2  \zeta
+ a^3 \dot{\zeta}^2 \zeta \right] - 2 \slrl ^2 \Mp^2 a^3 \,
\dot{\zeta} \left(\partial_i \partial^{-2}\dot{\zeta}\right)
\left(\partial_i \zeta\right)
- \frac{1}{2} \varepsilon_{1}^3 \Mp^2 a^3 \dot{\zeta}^2 \zeta
\nonumber \\ &
+2 \slrl  \Mp^4 a^3 \dot{\zeta} \zeta^2 \frac{\exd }{\exd  t}
\left( \frac{\ddot{\phi}}{2\dot{\phi}H}
+ \frac{\dot{\phi}^2}{4 H^2 \Mp^2} \right)
+ \frac{1}{2} \slrl ^3 \Mp^2 a^3 \left(\partial_{i}\partial_{j}
\partial^{-2} \dot{\zeta} \right)
\left(\partial_{i}\partial_{j}  \partial^{-2} \dot{\zeta} \right)
\zeta  \Biggr\rbrace \, .
\end{align}
This way of writing the cubic action is organized in increasing powers
of the slow-roll parameter, with the first line containing the
dominant terms and the rest being subdominant. Although the first line
is naively $\mathcal{O}(\slrl ^2)$ the quadratic
action~\eqref{freescalaraction_App} shows that correlations of $\zeta$
are themselves enhanced by slow-roll parameters. For this reason
slow-roll behaviour is easier to read when the action is expressed in
terms of $v$, and shows that the leading term of
\pref{allscalarcubics} is actually $\cO(\sqrt{\slrl })$.  

\section{Changing $\zeta \to v$ while tracking powers of $z$}
 
This appendix tracks how the dependence of the eigenvalues of the Lindblad equation change as one performs a change of variables from $v$ and $p$ to $\zeta$ and $\mfp$. This can matter when keeping track of powers of $z = - k\eta$ because the change of variables itself also becomes singular in the limit $z \to 0$. 

The Lindblad equation has the general form given in \pref{eq:lindbladlikeform}, 
\begin{equation}
\label{Appeq:lindbladlikeform}
\partial_{\eta} \varrho = -i\Bigl[ \scrH_{\rm eff}(\eta), \varrho(\eta)\Bigr] +\sum_{r,s=1}^2 h_{s r}\left[\mathcal{O}_{s}\varrho \, \mathcal{O}^{\dag}_{r}-\frac{1}{2}\left\{\mathcal{O}^{\dag}_{r}\mathcal{O}_{s},\varrho\right\}\right] \,,
\end{equation}
and we ask how the matrix of couplings $h_{rs}$ behaves under a change of field variables. These normally transform as
\be
  h_{rs} \to \mfh_{rs} = \Bigl( S^\ssT h S \Bigr)_{rs}
\ee
where $S_{rs}$ is the Jacobian of the field redefinition under which the Lindblad operators transform as $O_r = S_{rs} O_s$.

In our applications we are interested in a coupling matrix that has the schematic form
\be
    h = \lambda \left( \begin{matrix} 1 & \varepsilon \cr \varepsilon & 0 \end{matrix} \right)  \,,
\ee 
where $\varepsilon$ is a small quantity and $\lambda$ can be singular as $\varepsilon \to 0$. In the main text we found $\lambda \propto z^{-6}$ and $\varepsilon \propto z$ in the super-Hubble regime. The zero entry at bottom right is not strictly zero, but is more suppressed by powers of $\varepsilon$. Under a redefinition of the form
\be \label{AppSform}
    S = \left( \begin{matrix} a & b \cr c & d \end{matrix} \right)    \,,
\ee 
this coupling matrix becomes
\be
    \mfh =\lambda \left( \begin{matrix} a & c \cr b & d \end{matrix} \right)   \left( \begin{matrix} 1 & \varepsilon \cr \varepsilon & 0 \end{matrix} \right)
     \left( \begin{matrix} a & b \cr c & d \end{matrix} \right)    =\lambda \left[ \left( \begin{matrix} a^2 & ab \cr ab & b^2 \end{matrix} \right)   + \varepsilon \left( \begin{matrix} 2ac & ad+bc \cr ad+bc & 2bd \end{matrix} \right) \right] \,.
\ee 

Consider now a change of variables from $\{v , p\}$ to $\{ \zeta, \zeta' \}$ with 
\be \label{Appvzeta}
    \left( \begin{matrix} \zeta \cr \zeta'/k \end{matrix} \right)   = \alpha \left( \begin{matrix} 1 & 0 \cr c & 1 \end{matrix} \right)  \left( \begin{matrix} v \cr p/k \end{matrix} \right)   \,,
\ee 
where 
\be
   \alpha = \frac{k}{a(\eta) M_p \sqrt{2\epsilon_1}} \sim \cO(z) \,,
\ee
with the factor $k$ chosen to make the result dimensionless and $c = 1/(k\eta) \sim 1/z$.  Then 
\be
    \mfh  =\lambda \alpha^2 \left[ \left( \begin{matrix} 1 & 0 \cr 0 & 0 \end{matrix} \right)   + \varepsilon  \left( \begin{matrix} 2c & 1 \cr 1 & 0 \end{matrix} \right) \right] = \lambda \alpha^2  \left( \begin{matrix} 1+ 2\mfc & \varepsilon \cr \varepsilon & 0 \end{matrix} \right)   \,,
\ee 
where the last equality writes $c \propto z^{-1} = \mfc/\varepsilon$ with $\mfc$ being $\cO(1)$. All other terms are suppressed by at least one power of $\varepsilon$. This shows how the momentum transformation property changes the order-unity part of the eigenvalue, on top of the overall factor of $\alpha^2$.

At face value we know that any successful controlled derivation of the Lindblad equation cannot produce the couplings as above with a bottom-right entry that is smaller than order $\varepsilon^2$, because if so the determinant of both $h$ and $\mfh$ is negative (indicating the presence of a negative eigenvalue, and so also unitarity violating effects). Suppose then the lower-right entry of $h_{mn}$ is nonzero, with
\be
    h = \lambda \left( \begin{matrix} 1 & \varepsilon \cr \varepsilon & r \varepsilon^2 \end{matrix} \right) \,,
\ee 
where $r$ is order unity. The eigenvalues of $h_{mn}$ are
\bea
   \lambda_+ &=& \tfrac12 \lambda \Bigl[ 1 + \varepsilon^2 r  + \sqrt{1 + 2\varepsilon^2(2- r)+ \varepsilon^4 r^2}\Bigr] \simeq \lambda \Bigl[ 1 + \varepsilon^2 + \cO(\varepsilon^4) \Bigr] \cr
   \lambda_- &=& \tfrac12 \lambda \Bigl[ 1 + \varepsilon^2 r  - \sqrt{1 + 2\varepsilon^2(2- r)+ \varepsilon^4 r^2}\Bigr] \simeq \lambda \Bigl[ \left( r - 1 \right) \varepsilon^2 + \cO(\varepsilon^4) \Bigr]  \,,
\eea 
which show that positivity requires $r \geq 1$.

In this case a rotation of the form \pref{AppSform} gives
\be
    \mfh =\left( \begin{matrix} a & c \cr b & d \end{matrix} \right)   \left( \begin{matrix} \lambda & \varepsilon \cr \varepsilon & r \varepsilon^2 \end{matrix} \right)
     \left( \begin{matrix} a & b \cr c & d \end{matrix} \right)    =\lambda \left[ \left( \begin{matrix} a^2 & ab \cr ab & b^2 \end{matrix} \right)   + \varepsilon \left( \begin{matrix} 2ac & ad+bc \cr ad+bc & 2bd \end{matrix} \right)   + \varepsilon^2 \left( \begin{matrix} c^2 r & cdr \cr cdr & d^2 r \end{matrix} \right)  \right]\,,
\ee 
and so for the transformation \pref{Appvzeta} relating $\{v , p\}$ and $\{ \zeta, \zeta' \}$ we get
\be
    \mfh  =\lambda  \alpha^2  \left( \begin{matrix} 1 + 2\mfc + \mfc^2 r && \varepsilon(1 + \mfc r) \cr \varepsilon(1 + \mfc r) && \varepsilon^2 r \end{matrix} \right)   \,,
\ee 
with eigenvalues 
\be
   \mfl_+  \simeq   \lambda  \Bigl[ 1+2\mfc + \mfc^2 r + \varepsilon^2 + \cO(\varepsilon^4) \Bigr] \quad \hbox{and} \quad
   \mfl_- \simeq \lambda \Bigl[ \left( r - 1 \right) \varepsilon^2 + \cO(\varepsilon^4) \Bigr] \,,
\ee 
up to entries that are subdominant in $\varepsilon$. In particular the leading contribution to $\mfl_+$ receives an order-unity correction involving $r$. 

The upshot of this calculation is that it can be wrong to completely neglect the off-diagonal and bottom-right entry of the coupling matrix even if these are suppressed by powers of $\varepsilon \propto z$. This is why we keep the off-diagonal contributions when computing the purity, even though naively the $vv$ coupling of the Lindblad evolution is dominant. 

\section{Comparison to \href{https://arxiv.org/abs/2211.11046}{arXiv:2211.11046}}  
\label{App:Comparison}

This appendix makes a detailed comparison of the current calculation and the earlier one given in \cite{Burgess:2022nwu}. The goal is to clarify the differences between the two results, and how these are related to the assumptions made in each case. 

The calculation done in \cite{Burgess:2022nwu} differs in two important ways from the current framework. The main difference is the omission of time-derivative terms, which corresponds to omitting the interactions \pref{eq:B:def2} through \pref{eq:B:def4} and dropping the $p_\env^2$ contribution in eq.~\pref{eq:B:def1}. This was done for simplicity but rationalized after the fact by the observation that the most important contributions to decoherence of super-Hubble system modes turned out to involve environmental modes who were also super-Hubble. This required choosing $z_{\rm in} \ll 1$ and for this reason in this appendix we compare to our current results in this same small-$z_{\rm in}$ limit.

This paper and that one both find the dominant contribution to be given by the Markovian $n=0$ term, but in \cite{Burgess:2022nwu} this is further approximated by approximating $v_{-\bmk}(\eta') \simeq v_{-\bmk}(\eta)$, which amounts to keeping only the leading contribution in the Taylor expansions of Eqs.~\pref{eq:etaprimeToetav}. 

To make this comparison explicit, we rederive here the Markovian Nakajima-Zwanzig equation for the Mukhanov-Sasaki degree of freedom $v_{\bmk}(\eta)$ under the assumption that the only relevant coupling is the one used in \cite{Burgess:2022nwu}. The result reads
\begin{equation}
\partial_{\eta}\varrho_{\bmk}(\eta) = -\mathcal{J}^{v v}_{k 0}(\eta;\eta_{in})\Bigl[\zeta_{\bmk}(\eta),\ v_{-\bmk}(\eta)\ \varrho_\bmk(\eta)\Bigr] +\cdots,
\end{equation}
where 
\begin{equation}
\mathcal{J}^{v v}_{k 0}(\eta;\eta_{\in}) = {G}(\eta)  \int_{\eta_{\in}}^{\eta} \exd\eta^{\prime}\  {G}(\eta^{\prime}) \, \mathcal{T}^{vv}_{k}(\eta^{\prime},\eta),
\end{equation}
and
\begin{equation}
\mathcal{T}^{vv}_{k}(\eta^{\prime},\eta)= \scrW_k (\eta^{\prime}, \eta) \Bigl\langle  \cE^{v}(\eta) \, \cE^{v}(\eta^{\prime})\Bigr\rangle.
\end{equation}
Here $\cE^{v}(\eta)$ denotes the Fourier transform of $\left(\partial v\right)^2$ and $\scrW_k (\eta^{\prime}, \eta)$ is given in eq.\pref{eq:etaprimeToeta2v}. 

Now comes the main point. In \cite{Burgess:2022nwu} we replaced $\scrW_k (\eta^{\prime}, \eta)$ with its coincident limit $\scrW_k (\eta, \eta) = 1$, on the grounds that the correlation function $ \left\langle  \cE^{v}(\eta) \, \cE^{v}(\eta^{\prime})\right\rangle$ is sharply enough peaked at $\eta^{\prime}=\eta$ only the coincidence limit matters. Using the Wronksian condition \pref{eq:etaprimeToetav} for the modes, this amounted to the replacement of $\mathcal{T}^{vv}_{k}(\eta^{\prime},\eta)$ by just $\left\langle \cE^{v}(\eta)  \, \cE^{v}(\eta^{\prime})\right\rangle$ which led to the replacement 
\begin{equation}
\mathcal{J}^{v v}_{k 0}(\eta;\eta_{\in}) \rightarrow  \mfF_k(\eta) := G(\eta)  \int_{\eta_{\in}}^{\eta} \exd\eta^{\prime}\  {G}(\eta^{\prime})\left\langle  \cE^{v}(\eta)  \cE^{v}(\eta^{\prime})\right\rangle, 
\end{equation}
which is what we called the Lindblad coefficient in \cite{Burgess:2022nwu}. 

The question is whether this replacement captures the correct leading approximation (putting aside the validity of the assumption that only the one interaction is important). The result found in \cite{Burgess:2022nwu} for the real part of $ \mathfrak{F}_k(z)$ in the regime $z \ll z_{\rm in} \ll 1$ was 
\begin{equation}  \label{cFkOld}
  \hbox{Re}[\mfF_k(z)] \simeq \frac{5\pi}{4z^2} + \cO(z^{-1}) \,,
\end{equation}
wherewe switch to $z=-k\eta$ (as in the main text). To test how good this approximation is we can use the techniques developed in the main text to calculate the result without assuming the coincident limit for $\scrW_k(\eta', \eta)$. This gives the small-$z$ expansion
\begin{eqnarray}
\label{eq:comparisonA}
\hbox{Re} \left[\mathcal{J}^{v v}_{k 0}(z;z_{\in}) \right]&\simeq&\frac{1}{z^2} \left( -\frac{1}{2\epsilon^3}+\frac{1}{2\epsilon}\right) -\frac{1}{z} \left(\kappa + \frac{1}{4}+\frac{2}{3\kappa} -\frac{1}{8\kappa^2}\right)  \\
&&\qquad\qquad\qquad +\frac{4}{3 z_{\in}} \left(\kappa + \frac{1}{4}+\frac{2}{3\kappa} -\frac{1}{8\kappa^2}\right) -\frac{1}{3 \epsilon}  -\frac{\pi}{4}+ \cdots \,,\nn
\end{eqnarray}
and evaluating \pref{eq:comparisonA} using the regularization conventions of this paper gives (for small $z$)
\begin{equation}\label{eq:comparisonF}
  \hbox{Re}[ \mathfrak{F}_k(z) ] \simeq \frac{1}{z^2} \left( \frac{5 \pi}{4}-\frac{1}{2\epsilon^3} \right) +\frac{1}{2z}  \left(\kappa + \frac{1}{4}+\frac{2}{3\kappa} -\frac{1}{8\kappa^2}\right)  \ln\frac{z}{z_{\in}} -\frac{5}{12 \epsilon} -\frac{23 \pi}{24}+ \cdots \,,
\end{equation}
where (as in the main text) $\kappa := {\kUV}/{k}  > 1$. The result \pref{eq:comparisonF} agrees with \pref{cFkOld} to the extent that the UV regularization scheme used in \cite{Burgess:2022nwu} essentially set any positive powers of $1/\epsilon$ to zero. 

But the expression \pref{eq:comparisonF} for $\mfF_k$ does {\it not} agree with expression \pref{eq:comparisonA}, even at leading order in $z$. This means that the $n \neq 0$ corrections to $\mfF_k(z)$ must also contribute to the leading $1/z^2$ term. These corrections are given by
\begin{equation}
\hbox{Re}\left( \mathfrak{F}^{(n)}_k(z)\right) := \int^{z_{\in}-z-i\varepsilon z_{\in}}_{-i \varepsilon z} dw\ z(z+w)\ w^n \left( \frac{\partial^n_w \mathcal{A}_k (z+w, z)}{n!}\right)_{w=0}  \cC_k(z, z+w), 
\end{equation}
where 
\begin{eqnarray}\label{eq:kernels}
\mathcal{A}_k (z^{\prime}, z) &=&-\frac{i}{2}\Bigl[ \tilde{u}_k^* (z^{\prime}) \partial_z \tilde{u}_k(z)-\tilde{u}_k (z^{\prime}) \partial_z \tilde{u}_k^*(z)\Bigr]  \\
\mathcal{B}_k (z^{\prime}, z) &=& -\frac{i}{2}\Bigl[ \tilde{u}_k^* (z^{\prime}) \tilde{u}_k(z)-\tilde{u}_k (z^{\prime})  \tilde{u}_k^*(z)\Bigr]  \,.
\end{eqnarray}
with 
\begin{equation}\label{tildeudef}
 \tilde u_k = \sqrt{2k} \, u_k = \left( 1 + \frac{i}{z} \right) e^{iz} \quad \hbox{and} \quad \scrC_{\bmk}(\eta , \eta') = k^5 \cC_{k}(z,z') \,.
\end{equation} 
where $\scrC_\bmk$ is defined in terms of the $\partial v(\eta, \hbox{\bf 0}) \, \partial v(\eta', \bmy)$ correlator $C_\ssB(\eta, \eta', \bmy)$ by
\begin{equation}
\label{CR_FT}
C_\ssB(\eta, \eta' ; \bm{y}) = \int \frac{\exd^{3} \bm{k}}{(2\pi)^{3/2}}
\mathscr{C}_{\bm{k}}(\eta,\eta') e^{i \bm{k} \cdot \bm{y} } \,.
\end{equation}
Evaluating these shows that the $1\slash z^2$ term receives contributions from out to the sixth moment $\hbox{Re}\left( \mathfrak{F}^{(6)}_k(z)\right)$, and summing these leads to the leading small-$z$ result:
\begin{equation}
\frac{1}{z^2} \left(- \frac{1}{2\epsilon^3}+\frac{1}{2\epsilon}\right)\,,
\end{equation}
in agreement with the leading term of \pref{eq:comparisonA}. 

The upshot is this: for the variable $v_\bmk$ the analog of \pref{eq:TCL2} is:
\begin{eqnarray}
\label{eq:TCL2dimlessandfactored}
\partial_{\eta} \varrho &=& -\sum_{n=0}^{\infty} \frac{1}{ n!} \sum_{\bmk } \left\{k^{2-n} \mathcal{A}^{(n)}_k(z;z_{\rm in}) \Bigl[v_{\bmk }(\eta), v_{-\bmk }(\eta) \, \partial_{\eta}^n \varrho(\eta) \Bigr]  \phantom{\int} \right. \\
&&\qquad\qquad \qquad \qquad\qquad\qquad \qquad \left. \phantom{\int} + k^{1-n} \mathcal{B}^{(n)}_k(z;z_{\rm in}) \Bigl[v_{\bmk }(\eta), p_{\bmk }(\eta) \, \partial_{\eta}^n \varrho(\eta) \Bigr]\right\} \,, \nn
\end{eqnarray}
where the convolution integrals are 
\begin{equation} \label{akn}
  \mathcal{A}^{(n)}_k(z; z_{\rm in}) = \int^{z_{in}-z-i\varepsilon z_{\rm in}}_{-i \varepsilon z} dw\ z(z+w)\ (-w)^n  \cC_k(z, z+w)\mathcal{A}_k (z+w, z) \,,
\end{equation}
and 
\begin{equation} \label{bkn}
 \mathcal{B}^{(n)}_k(z; z_{\rm in}) = \int^{z_{in}-z-i\varepsilon z_{\rm in}}_{-i \varepsilon z} dw\ z(z+w)\ (-w)^n  \cC_k(z, z+w)\, \mathcal{B}_k (z+w, z) \,,
\end{equation}
where the quantities appearing in the integrands are defined by eqs.~\pref{eq:kernels} through \pref{tildeudef}. 

Although for super-Hubble modes the integrand appearing in the evolution equation \pref{eq:TCL2dimlessandfactored} is peaked enough to justify a Markovian approximation involving no derivatives of $\varrho_\bmk$, this approximation is not well captured by dropping all but the leading power of $w$ in the expressions for $\cA_k(z',z)$ and $\cB_k(z',z)$. Although \cite{Burgess:2022nwu} checked that the first subleading power of $w$ was subdominant, this was misleading because it is accidentally small due to the absence of the first subdominant term in the full $w$ expansion 
\begin{eqnarray}
\label{eq:AkTaylorExp}
\mathcal{A}_k (z^{\prime}, z)&\simeq& 1+\left(-\frac{1}{2}+\frac{1}{z^2}\right) w^2-\frac{2}{3 z^3} w^3+\cdots \\
\label{eq:BkTaylorExp}
\mathcal{B}_k (z^{\prime}, z)&\simeq& -w+\left(\frac{1}{6}-\frac{1}{3 z^2}\right) w^3+\cdots \,. 
\end{eqnarray}

\section{Lindblad evolution}
\label{App:LindTransport}

We here derive the Lindblad equation used in the main text. We begin with $\mathrm{TCL}_{2}$ equation Eq.~(\ref{eq:TCL2a}), keeping only the $n=0$ terms so that
\bea
\partial_{\eta} \varrho_{\bmk }(\eta) &=& -  \mathcal{J}^{\zeta \zeta}_{k0}(\eta;\eta_{\mathrm{in}})\Bigl[\zeta_{\bmk}(\eta),\ \zeta_{-\bmk}(\eta)\ \varrho_{\bmk}(\eta)\Bigr] - \mathcal{J}^{\zeta \mfp}_{k0}(\eta;\eta_{\mathrm{in}})\Bigl[\zeta_{\bmk}(\eta),\ \mfp_{-\bmk}(\eta)\ \varrho_{\bmk}(\eta)\Bigr] \\
&&\;  - \mathcal{J}^{\mfp \zeta}_{k 0}(\eta;\eta_{\mathrm{in}})\Bigl[\mfp_{\bmk}(\eta),\ \zeta_{-\bmk}(\eta)\ \varrho_{\bmk}(\eta)\Bigr] - \mathcal{J}^{\mfp \mfp}_{k 0}(\eta;\eta_{\mathrm{in}})\Bigl[\mfp_{\bmk}(\eta),\ \mfp_{-\bmk}(\eta)\ \varrho_{\bmk}(\eta)\Bigr] + \mathrm{h.c.}\nn
\eea
and we will now take $\eta_{\mathrm{in}} \to - \infty$. For now ignoring the $\bmk$ dependence of operators (using the real/imagainary decomposition described in eq.~\pref{alphaRI}) we find that for terms with two of the same operators, we have the simple identities:
\bea
\ [ X,X\varrho] + [ \varrho X, X ] & = & \left[ X, [ X, \rho ] \right]  \ = \ - 2 \big( X \rho X - \tfrac{1}{2} \{ X^2 , \rho \} \big)  \\
\ [ X , X \varrho ] - [ \varrho X, X ] & = & [ X^2 , \rho ]
\eea
As well as:
\begin{eqnarray} 
- [ \zeta, \mfp \varrho ] - [ \varrho \mfp , \zeta ] - [ \mfp, \zeta \varrho ] - [ \varrho \zeta , \mfp ] & = & + 2  \big( \zeta \varrho \mfp - \tfrac{1}{2} \{ \mfp \zeta, \varrho \}  \big) + 2 \big( \mfp \varrho \zeta - \tfrac{1}{2} \{ \zeta \mfp, \varrho \} \big) \\
- [ \zeta, \mfp \varrho ] + [ \varrho \mfp , \zeta ] + [ \mfp, \zeta \varrho ] - [ \varrho \zeta , \mfp ] & = & - 2  \big( \zeta \varrho \mfp - \tfrac{1}{2} \{ \mfp \zeta, \varrho \}  \big) + 2 \big( \mfp \varrho \zeta - \tfrac{1}{2} \{ \zeta \mfp, \varrho \} \big) \\
- [ \zeta, \mfp \varrho ] - [ \varrho \mfp , \zeta ] + [ \mfp, \zeta \varrho ] + [ \varrho \zeta , \mfp ] & = & - \big[  [\zeta,\mfp], \varrho \big] \\
- [ \zeta, \mfp \varrho ] + [ \varrho \mfp , \zeta ] - [ \mfp, \zeta \varrho ] + [ \varrho \zeta , \mfp ] & = & - \big[  \{ \zeta, \mfp \}, \varrho \big] 
\end{eqnarray}
Treating $\mathrm{O}_{1} = \zeta$ and $\mathrm{O}_{2} = \mfp$ the above implies that the $\mathrm{TCL}_{2}$ equation gets put into the Lindblad form
\bea
\partial_{\eta} \varrho_{\bmk } & = & - i [ \scrH_{\rm eff} , \varrho_{\bmk} ] + \sum_{r,s = 1,2} \mathrm{h}_{rs} \Big( \mathrm{O}_{r} \varrho_{\bmk} \mathrm{O}_{s} - \tfrac{1}{2} \{ \mathrm{O}_{s} \mathrm{O}_{r}, \varrho_{\bmk}  \} \Big)
\eea
with
\begin{eqnarray}
\scrH_{\rm eff} = \mathrm{Im}[ \mathcal{J}^{\zeta \zeta}_{k0}] \zeta^2 - \mathrm{Im}[\mathcal{J}^{\zeta \mfp}_{k0} + \mathcal{J}^{\mfp \zeta}_{k0}]  \tfrac12 \{\zeta,\mfp \}  + \mathrm{Im}[\mathcal{J}^{\mfp \mfp}_{k0}] \mfp^2
\end{eqnarray} 
and
\begin{equation}
\left[ \begin{matrix} \mathrm{h}_{11} & \mathrm{h}_{12} \\ \mathrm{h}_{21} & \mathrm{h}_{22} \end{matrix} \right] = \left[ \begin{matrix} 2\mathrm{Re}[\mathcal{J}^{\zeta \zeta}_{k0} ] & \mathcal{J}^{\zeta \mfp \ast}_{k0} + \mathcal{J}^{\mfp \zeta}_{k 0}  \\ \mathcal{J}^{\zeta \mfp}_{k0} + \mathcal{J}^{\mfp \zeta \ast}_{k 0}  & 2 \mathrm{Re}[\mathcal{J}^{\mfp \mfp}_{k0} ] \end{matrix} \right] \ .
\end{equation}
Notice that the determinant of this matrix is
\bea
\det \left[ \begin{matrix} \mathrm{h}_{11} & \mathrm{h}_{12} \\ \mathrm{h}_{21} & \mathrm{h}_{22} \end{matrix} \right] & = & 4 \mathrm{Re}[\mathcal{J}^{\zeta \zeta}_{k0} ] \mathrm{Re}[\mathcal{J}^{\mfp \mfp}_{k0} ] - \mathrm{Re}[ \mathcal{J}^{\zeta \mfp \ast}_{k0} + \mathcal{J}^{\mfp \zeta}_{k 0}]^2 - \mathrm{Im}[ \mathcal{J}^{\zeta \mfp \ast}_{k0} - \mathcal{J}^{\mfp \zeta}_{k 0}]^2 
\eea
In the main body, we instead work with the rescaled dimensionless jump operators
\begin{equation}
\mathcal{O}_{1} = Z^{-1} \zeta \quad \mathrm{and}  \quad \mathcal{O}_{2} = Z \mfp \qquad \mathrm{with} \ Z =\frac{1}{\sqrt{2 \epsilon_1 k^3}} \frac{H}{M_p}  
\end{equation}
as well as the dimensionless coefficients defined in eq.~(\ref{curlyJdimless}), which turns the Lindblad equation into
\begin{equation}
\partial_{\eta} \varrho_{\bmk} = -i\Bigl[ \scrH_{\rm eff}(\eta), \varrho_{\bmk}(\eta)\Bigr] +\sum_{r,s=1}^2 h^{s r}_\bmk\left[\mathcal{O}_{\bmk,s}\varrho_{\bmk}\mathcal{O}^{\dag}_{\bmk,r}-\frac{1}{2}\left\{\mathcal{O}^{\dag}_{\bmk,r}\mathcal{O}_{\bmk,s},\varrho_{\bmk}\right\}\right] 
\end{equation}
where the Kossakowski matrix is now
\begin{equation}
\left[ \begin{matrix} h_{\bmk}^{11} & h_{\bmk}^{12} \\ h_{\bmk}^{21} & h_{\bmk}^{22} \end{matrix} \right] = \left[ \begin{matrix} 2Z^{2}\mathrm{Re}[\mathcal{J}^{\zeta \zeta}_{k0} ] & \mathcal{J}^{\zeta \mfp \ast}_{k0} + \mathcal{J}^{\mfp \zeta}_{k 0}  \\ \mathcal{J}^{\zeta \mfp}_{k0} + \mathcal{J}^{\mfp \zeta \ast}_{k 0}  & 2 Z^{-2} \mathrm{Re}[\mathcal{J}^{\mfp \mfp}_{k0} ] \end{matrix} \right] = \frac{\slrl k}{8}\left(\frac{H}{\Mp}\right)^2\left(\begin{array}{cc}{2 \rm Re}[ \widetilde{\cJ}^{\zeta\zeta}_{k0} ] & \widetilde{\cJ}^{\zeta\mfp}_{k0} + \widetilde{\cJ}^{\mfp\zeta*}_{k0} \\ \widetilde{\cJ}^{\mfp\zeta}_{k0} +\widetilde{\cJ}^{\zeta\mfp *}_{k0} & 2 {\rm Re}[\widetilde{\cJ}^{\mfp\mfp}_{k0}] \end{array}\right)
\end{equation}
as quoted in (\ref{eq:decohmatrixh}).

\bibliographystyle{JHEP}
\bibliography{dSDecoherenceTwo.bib}

\end{document}